\documentclass[useAMS,usegraphicx,usenatbib]{mn2e}

\usepackage{amssymb}

\title{The Radio - X--ray relation as a star formation indicator: Results from the VLA--E-CDFS Survey}
\author[S. Vattakunnel et al.]{S. Vattakunnel$^{1}$, P. Tozzi$^{2}$, F. Matteucci$^{1,2}$, P. Padovani$^{3}$, N. Miller$^{4}$,
\newauthor M. Bonzini$^{3}$, V. Mainieri$^{3}$, M. Paolillo$^{5}$, L. Vincoletto$^{1}$, W.N. Brandt$^{6,7}$,
\newauthor B. Luo$^{6,7}$, K.I. Kellermann$^{8}$, Y.Q. Xue$^{6,7}$\\
$^{1}$Dipartimento di Fisica Universit\`{a} di Trieste, piazzale Europa 1, I--34127, Trieste, Italy\\
$^{2}$INAF Osservatorio Astronomico di Trieste, via G.B. Tiepolo 11, I--34143, Trieste, Italy\\
$^{3}$European Southern Observatory, Karl-Schwarzschild-Str. 2, D-85748 Garching bei M\"{u}nchen, Germany\\
$^{4}$Department of Astronomy, University of Maryland, College Park, MD 20742, USA\\
$^{5}$Dipartimento de Scienze Fisiche, Universit\`{a} di Napoli, Via Cinthia, 80126 Napoli, Italy\\
$^{6}$Department of Astronomy and Astrophysics, Pennsylvania State University, University Park, PA 16802, USA\\
$^{7}$Institute for Gravitation and the Cosmos, Pennsylvania State University, University Park, PA 16802, USA\\
$^{8}$National Radio Astronomy Observatory, 520 Edgemont Road, Charlottesville, VA 22903-2475, USA}

\begin{document}

\date{Received 25 July 2011; Accepted 11 November 2011}

\pagerange{\pageref{firstpage}--\pageref{lastpage}} \pubyear{2011}

\maketitle

\label{firstpage}

\begin{abstract}
In order to trace the instantaneous star formation rate at high
redshift, and hence help understanding the relation between the
different emission mechanisms related to star formation, we combine
the recent 4 Ms {\it Chandra} X-ray data and the deep VLA radio data
in the Extended {\it Chandra} Deep Field South region. We find 268
sources detected both in the X-ray and radio band.  The availability
of redshifts for $\sim$95\% of the sources in our sample allows us to
derive reliable luminosity estimates and the intrinsic properties from
X-ray analysis for the majority of the objects. With the aim of
selecting sources powered by star formation in both bands, we adopt
classification criteria based on X-ray and radio data, exploiting the
X-ray spectral features and time variability, taking advantage of
observations scattered across more than ten years.  We identify
43 objects consistent with being powered by star formation.
We also add another 111 and 70 star forming candidates detected
only in the radio or X-ray band, respectively.  We find a clear linear
correlation between radio and X-ray luminosity in star forming
galaxies over three orders of magnitude and up to $z \sim
1.5$. We also measure a significant scatter of the order of 0.4
  dex, higher than that observed at low redshift, implying an
intrinsic scatter component.  The correlation is consistent with that
measured locally, and no evolution with redshift is observed.  Using a
locally calibrated relation between the SFR and the radio luminosity,
we investigate the $L_{X}(2-10keV)$-SFR relation at high redshift.
The comparison of the star formation rate measured in our sample with
some theoretical models for the Milky Way and M31, two typical spiral
galaxies, indicates that, with current data, we can trace typical
  spirals only at $z\leq 0.2$, and strong starburst galaxies with
  star-formation rates as high as $\sim 100 M_\odot yr^{-1}$, up to $z\sim 1.5$.

\end{abstract}


\section{Introduction}
The understanding of when and how galaxies formed is one of the most
difficult and relevant issues in astrophysics.  It is therefore
important to understand the key elements of the evolutionary history
of galaxies and the physical processes which led to the complexity of
the Universe as we see it now, with galaxies of different
morphologies, ages and chemical compositions.  The wide ranges in
stellar content and star formation activity present in the locally
observed Hubble sequence are vital clues for understanding galaxy
evolution, where many physical processes, from mergers to shocks, from
accretion to winds, play an important role.  However, even with the
most advanced telescopes, it is not possible to resolve individual
stars in any but the closest galaxies, so we have to rely on
integrated light measurements in order to trace young stellar
populations and, therefore, star formation rate (SFR), i.e. the rate
at which the gas is turned into stars.

Spectral synthesis modelling is at the basis of several methods
  of SFR measurements.  Observing galaxy spectra along the Hubble
sequence shows a gradual change in the spectral shape, which is mainly
dictated by the ratio of early to late-type stars.  Optical
luminosities and colors are modelled using stellar evolution tracks
and atmosphere models, assuming an initial mass function
(IMF). However, the estimate of SFR obtained in this way is relatively
imprecise. There are two major sources of uncertainties. One is due to
errors associated with the extinction correction; the other one is due
to uncertainties in the modelling, namely the shape of the IMF and the
age-metallicity degeneracy.

The best estimates of the immediate SFR can be derived by observations
at wavelengths where the integrated spectrum is dominated by young and
massive stars. At the high end of the IMF the SFR scales linearly with
the measured luminosity.  Whenever a significant dust component is
present, the light emitted in the ultra-violet (UV) and optical bands
by young stars is absorbed by interstellar dust and re-emitted in the
far-infrared (FIR).  Therefore, we can complement the UV observations
with the FIR luminosity to measure the instantaneous SFR.  However,
the efficient use of FIR luminosity as a SFR tracer depends on the
contribution of young stars' UV emission to the heating of the dust,
and on the optical depth of the dust.  This contribution is generally
difficult to estimate \citep{buat96,calzetti92}.

Some other important direct tracers of star formation are nebular
emission lines, mainly the H$\alpha $ line, which re-emit the
integrated stellar luminosity of the young population; i.e., OB stars
which are usually surrounded by H II regions \citep{kennicutt83}.
However, this method is sensitive to extinction which affects the
H$\alpha $ observed fluxes, not to mention the possible contamination
by non thermal optical nuclear emission which is difficult to
disentangle from emission solely due to star formation.
Forbidden lines can provide a valid alternative. The strongest emission 
feature in the blue is the [OII]
$\lambda$3723 line doublet. Even if not directly coupled with the
ionizing luminosity, it can be calibrated empirically as a star
formation tracer \citep{gallagher89,kennicutt92a}, but is
less precise than H$\alpha $.

The UV, optical and IR observational windows still remain the most
popular way to measure the star formation rate in galaxies over a wide
range of redshifts.  Great efforts are made in order to overcome the
disadvantages which affect these bands, from the dust absorption to
the handling of the several processes which contribute to the
emission.  An alternative and independent way to explore the cosmic
star formation history is provided by two star-formation tracers which
are unaffected by dust: radio and X--ray emission.  The SF related
emission at radio frequencies is associated with non-thermal processes
predominantly due to synchrotron radiation from supernova explosions
and thermal bremsstrahlung from HII regions.  Relations between the
SFR and radio luminosity have been explored by many authors (e.g.,
\citealt{condon92,yun01,bell03,schmitt06}).  These observables are
sensitive to the number of the most massive stars and provide a direct
measure of the instantaneous SF above a certain mass. We also
  remind that a very tight correlation is observed between the radio
  and IR emission in SF galaxies at low (see Bell 2003) and high
  redshift \citep{mao11}.  It should be noted, however, that there is
no clear theoretical connection between radio luminosity and SFR, and
the widely used relations are derived empirically.

SF related X--ray emission is instead due to High and Low Mass X-ray
Binaries (HMXB and LMXB), young supernova remnants, and hot plasma
associated with star-forming regions (e.g.,
\citealt{fabbiano89a,fabbiano94,fabbiano06}).  LMXB and HMXB mainly
differ in the mass of the accreting neutron star companion, or black
hole, and have very different evolutionary time-scales. Hence, only
short-living HMXBs, whose evolutionary time scale does not exceed
$\approx 10^7$ yrs, can be used as a tracer for the instantaneous star
formation activity in the host galaxy, while LMXBs are connected to
the past star formation and the total stellar content
\citep{grimm03}. In the soft band (0.5-2 keV) the X--ray binary
emission can be confused with thermal emission from the surrounding
hot gas not directly related to SF, but rather associated to the
gravitationally-heated gas residing in the host galaxy halo. In the
hard band (2-10 keV), where the dust is almost transparent to the
X-rays, the SF related emission is dominated by X-ray binaries and
therefore is a more robust SF indicator.  In general, X-ray
  emission is considered a good estimator of SFR also thanks to the
  observed correlations. For example, \citet{fabbiano84b} first
  investigated the relationship between X--ray and radio/optical
  luminosities in a nearby sample of spiral galaxies. The most studied
  correlation is between X-ray and radio luminosities.  Several works
  established a clear X-ray/radio correlation for star forming
  galaxies \citep{bauer02,ranalli03,grimm03,gilfanov04}. In particular
  \citet{ranalli03} studied the local correlation between soft/hard
  X--ray and radio/infrared luminosities. Using the relations found by
  \citet{condon92} and \citet{kennicutt98} relating SFR to radio and
  infrared emission, they calibrated the X--ray soft and hard
  luminosities as SFR indicators.  More recently \citet{persic07}
  analysed the correlation between FIR estimated SFR and the
  collective emission of X-ray point sources (assumed to be dominated
  by HMXBs) in a local sample of star forming galaxies (SFG), finding
  similar results to \citet{ranalli03}.  \citet{grimm03} studied the
  relation between star formation and the population of high-mass
  X--ray binaries in nearby star forming galaxies. They found that the
  X-ray luminosity is directly proportional to SFR at sufficiently
  high values of star formation rate, while at lower values the
  relation becomes non-linear.

  At high redshift, the existence of an X-ray/radio correlation for
  star forming galaxies is more uncertain.  \citet{bauer02}
  investigated at high redshift the X-ray radio luminosity correlation on
  the basis of 20 emission line galaxies identified in the Chandra
  Deep Field North (CDFN), finding a linear relation in agreement with
  local estimates and concluding that the X-ray emission can be used
  as a SFR indicator also at high z.  The robustness of the X-ray/radio relation for 
  SFG at high redshift has been found also with stacking of UV selected 
  galaxies at $1.5 < z < 3$ in the CDFN \citep{reddy04}.  On the contrary,
  \citet{barger07} argued against the existence of such a correlation
  on the basis of the VLA radio and 2 Ms X-ray data in the CDFN for a
  spectroscopically identified sample of star forming
  galaxies. Recently \citet{symeonidis11} investigated the X-ray/IR
  correlation for SFG at redshift 1 and found a non-evolving linear
  relation consistent with local estimates, concluding that the X-ray
  luminosity can be used as a tracer of star formation.

The purpose of this paper is to explore further this aspect and derive
the SFRs in high z galaxies from the deepest X--ray and radio
observations available to date using radio and X--ray surveys of the
{\it Chandra} Deep Field South (CDFS). The CDFS is located in a sky
region with low Galactic absorption, and it has been observed by deep,
multiwavelength surveys. The radio data consist of high-resolution 1.4
GHz (21 cm) imaging with the Very Large Array (VLA) across the full
Extended {\it Chandra} Deep Field South (E--CDFS). The X--ray data
consist of the currently deepest X--ray survey, with an exposure time
of about 4 Ms. This work is an extension of the work done with the
earlier observations of the X--ray and radio emission from the CDFS
(\citealt{kellermann08,mainieri08,tozzi09,padovani09,padovani11},
hereafter Papers I-V). With respect to previous papers, we make use of
deeper data and therefore of much larger source catalogs.  The main
consequence is that we do not have the optical and IR identification
for most of the newly detected sources in the radio and X-ray data.
Therefore, we identify SFGs on the basis of X-ray and radio emissions
alone.  Our goal is to evaluate the L$_R$-L$_X$ relation for star
forming galaxies at high redshift, estimate the SFR in the
star-forming galaxies in our sample, and finally compare our results
to some evolution models of SF galaxies. 

The paper is structured as follows: in Section \ref{sec:data} we
describe the radio and X--ray data and identify the X-ray counterparts
of all our radio sources. In Section \ref{sec:match} we perform a
photometric and spectroscopic analysis of X--ray counterparts,
discussing the different properties of radio sources. In Section 4 we
describe our selection criteria to identify SF galaxies and classify
each object as AGN or SFG on the basis of their radio and X-ray
properties. In Section 5 we derive the $L_X-L_R$ relation for SF
galaxies at high redshift and compare it to local estimates.  In
Section \ref{sec:SFR} we evaluate the SFR of selected objects, compare
them to theoretical evolution models, and discuss the prospects for
future surveys.  We summarize our results in Section \ref{sec:end}.
Luminosities are computed using the 7 years WMAP cosmology
($\Omega_{\Lambda} =0.73 $, $\Omega_m =0.27$, and $H_0 = 70.4 $ km
s$^{-1}$ Mpc$^{-1}$, see \citealt{komatsu11}).

\section{The data}\label{sec:data}

The {\it Chandra} Deep Field South is one of the most important deep
fields studied with multiwavelength observations. Several programs
based on the CDFS dataset have proved very useful to investigate the
complexity of the process of galaxy formation and evolution (see
\citealt{brandt05}).  The X-ray dataset originally consisted of
several {\it Chandra} ACIS-I observations for a total of about 1 Ms
(hereafter 1Ms observation) on a field of about 16' $\times$ 16'
\citep{giacconi02}, later extended to a field of 30' $\times$ 30'
centered on the original aimpoint with four flanking fields with an
exposure of 250 ks each (the Extended CDFS, hereafter E--CDFS,
\citealt{lehmer05}). Additional {\it Chandra} time was added in the
CDFS to reach a total of $\sim$2 Ms exposure in \citet{luo08}.  In
2010 the CDFS reached a total of 4 Ms and became the deepest X-ray
survey ever performed. The catalog of X-ray sources is presented in
\citet{xue11}. In this paper we use this data set for the spectral
analysis.

The VLA observation of the E--CDFS at 1.4GHz and 5GHz is presented in
Kellerman et al. (2008, Paper I) and is used in Papers I-V.  Later, new
data have been acquired at 1.4 GHz and a new deeper radio source
catalog has been obtained \citep{miller08}.  Important data in the
CDFS at other wavelengths are part of the Great Observatories Origins
Deep Survey (GOODS, see \citealt{dickinson03}), COMBO-17
(\citealt{wolf04}, \citealt{wolf08} the Hubble Ultra Deep Field (HUDF,
\citealt{beckwith06}), GEMS \citep{rix04}, the infrared Spitzer SIMPLE
\citep{damen11} and
FIDEL \footnote{http://irsa.ipac.caltech.edu/data/SPITZER/FIDEL/ }
surveys, the ultraviolet GALEX Ultra-Deep Imaging Survey and Deep
Spectroscopic Survey \citep{martin05}.  Many X-ray sources have been
already identified in the optical, IR and FIR data \citep{xue11}.  The
source identification and classification of the radio sources based
also on optical and infrared bands will be discussed in a subsequent
paper (Bonzini et al., in preparation)\nocite{bonzini11}.  In this paper we will
focus only on the X-ray and radio data, except for the spectroscopic
and photometric redshifts, and the X-ray to optical flux
ratio. In this section we describe in detail the X-ray and radio
datasets.

\subsection{X--ray observations}\label{sec:xdata}

In the E--CDFS area, we have two sets of X--ray data obtained with
{\it Chandra}. The most important is a 4 Ms exposure observation
resulting from the coaddition of 54 individual {\it Chandra} ACIS--I
exposures from October 1999 to July 2010 with centers spaced within a
few arcsec from $\alpha=$3:32:28.80, $\delta=-$27:48:23 (J2000).  To
ensure a uniform data reduction, we follow the same procedure adopted
in previous work (see \citealt{rosati02,tozzi06}) using the most
recent release of {\it Chandra} calibration files (CALDB 4.4).  The 4
Ms observations reach on-axis flux limits of $\approx 9.1 \times
10^{-18}$ and $\approx 5.5 \times 10^{-17}$ ergs cm$^{-2}$ s$^{-1}$
for the soft [0.5-2.0 keV] and hard [2-8 keV] bands, respectively (Xue
et al. 2011).  The 4 Ms main X-ray source catalog by Xue et al. (2011)
includes 740 X-ray sources and is used as the initial source input
list. We removed 8 uncertain objects because we do not have positive
aperture photometry for them in our data.  The catalog we use,
therefore, contains 732 sources.

The X-ray spectrum of each source is extracted from the total merged
event file. The size of the extraction region depends on the off-axis
angle between the source and the average aimpoint, as described in
\citet{giacconi01}.  The only difference in this new reduction is that
the minimum extraction radius is set to 3 arcsec as opposed to a
minimum of 5 arcsec.  This will further improve the signal-to-noise
ratio (S/N) for the faint sources lying in the central regions, where
the spatial resolution is about 1 arcsec.  The adopted extraction
radius is given by $R_{S}=2.8 \times FWHM \, \, arcsec$, where the
$FMHM=\sum_{i=0,2}a_{i} \theta^{i}$, with
$a_{i}=\{0.678,-0.0405,0.0535\}$. Here $\theta$ is the off-axis angle
in arcmin with respect to the aimpoint of the CDFS. The factor 2.8 for
the extraction radius has been chosen after a careful analysis of the
source photometry, by varying the factor in the range 1.0 - 4.0 and
inspecting the S/N of bright sources. In the cases where the
extraction regions of two nearby sources significantly overlap, we
resize them by hand until the regions are not overlapping.  In few
cases, for very bright sources, we also adopt a larger extraction
radius when, upon visual inspection, the wings of the emission
exceed the nominal $R_S$ value.

In order to create the response files for the spectral analysis, we
adopt the following procedure.  We create the response matrix files
and the ancillary response files for each exposure using the {\tt
  CIAO} software (version 4.3).  Since each source is typically
detected only in a subset of exposures, we compute the response files
whenever we find at least two counts in the extraction region in the
0.5-7 keV energy range\footnote{The 0.5-7 keV energy range is used to
  detect sources and to produce calibration files, while we use the
  full 0.5-10 keV energy band for the spectral analysis.}, and with an
average exposure larger than 30\% of the maximum value at the aimpoint
(this last condition is to avoid computing the files for sources whose
extraction region only partially overlaps with the edges of the CCD).
The final rmf and arf files are obtained by summing and weighting the
individual files according to the number of counts detected within the
extraction radius in each exposure.  In this way, we obtain the best
possible calibration file for each source, reflecting the complexity
of the CDFS exposures.

A further X-ray data set is the shallower $\sim 250$ ks coverage of
the square region of $0.28$ deg$^2$ centered on the 4 Ms field.
The E--CDFS consists of 9 observations of 4 partially overlapping
quadrants, 2 for each of the first three fields and 3 for the last one.
We applied the same data reduction procedure used for the
CDFS-4 Ms. The extraction regions are obtained with the same formula,
but in this case the off-axis angle is the angle between the source
and the aimpoint of each quadrant.  The original catalog by \citet{lehmer05} 
contains 762 sources.  We remove 9 detections because we do not 
have positive aperture photometry for them in our data reduction. 
The catalog we use, therefore, contains 753 sources.

The two data sets are treated separately, since it is not convenient
to add them due to the large differences in the point spread function
in the overlapping areas. In particular, the E--CDFS data have a much
better spatial resolution in the regions close to the edges of the
CDFS-4Ms image, due the broadening of the
PSF in the CDFS-4Ms data at large off-axis angle. Therefore we use
separately the two catalogs (the E-CDFS and the CDFS--4Ms), and then
merge {\sl a posteriori} the information obtained for the sources
detected in both data-sets. The total number of unique X-ray sources 
identified in the CDFS-4 Ms or E-CDFS is 1352.

\subsection{The radio observations}\label{sec:rdata}

We consider the data from the new VLA program which provides deep,
high resolution 1.4 GHz imaging across the full E--CDFS, consisting of
a six-pointing mosaic of 240 h spanning 48 days of individual 5 h
observations \citep{miller08}. The image covers a region of 34'.1
$\times$ 34'.1 of the full E--CDFS at a rms sensitivity of 7.5 $\mu$Jy
per 2.8'' $\times$ 1.6'' beam, with the rms reaching 7.2 $\mu$Jy per
beam in the central 30'. A S/N image is obtained by dividing the final
mosaic image by the rms sensitivity image. Then the sources are
identified as peaks in this S/N image.  Each detection is fitted with
a Gaussian to measure its position and extension.  The catalog
published by \citet{miller08} is based on a conservative detection
threshold at 7 $\sigma.$ Here we use two catalogs: a very deep one
including all the sources with S/N $>$ 4 (1571 sources) is used when
cross matching with X-ray detections. A more conservative catalog at
S/N $>$ 5 is used to investigate the properties of radio sources
without X-ray counterparts. This catalog includes 940 sources
  which shrinks to 879 single sources after identifying the multiple
  components (see Miller et al. 2011, in
  preparation)\nocite{miller11}.  The typical source positional
error has an average value of 0.2'' across the field of view, where a
0.1" portion of the error budget is constant across the full area and
the rest comes from noise in Gaussian fitting. For bright sources with
high SNR the error is only 0.1''.

We use also the radio data collected in 1999-2001 and presented in
Papers I-V.  The whole area of the E-CDFS has been observed with the
NRAO Very Large Array (VLA) for 50 h at 1.4 GHz mostly in the BnA
configuration in 1999 and February 2001, and for 32 h at 5 GHz mostly
in the C and CnB configurations in 2001.  The effective angular
resolution is 3.5'' and the minimum {\sl rms} noise is as low as 8.5
$\mu$Jy per beam at both 1.4 GHz and 5 GHz. Given the lower spatial
resolution and the lower sensitivity, we use these data only to
investigate variability in the 1.4 GHz band, and to obtain spectral
information for the subset of sources detected in the 5 GHz band.

\subsection{The optical data}

The E-CDFS area has been targeted by a large number of spectroscopic
surveys. For the X-ray sources we use the spectroscopic redshift
published in \citet{xue11}.  We also used the redshifts published
recently for the counterparts of {\it Chandra} sources in the same
field \citep{treister09,silverman10} and we include the dedicated
spectroscopic follow-up of the VLA sources in the E-CDFS performed
with the VIMOS spectrograph at the VLT (Bonzini et al., in preparation).  
We find
a total of 706 spectroscopic redshifts for unique X-ray sources in the
CDFS or E-CDFS fields. For the radio sources we collect the redshifts
from the publicly available catalogues: the follow-up campaign of the
X-ray sources in the CDFS \citep{szokoly04}, the FORS-2/GOODS program
\citep{vanzella05, vanzella06, vanzella08}, the VIMOS/GOODS program
\citep{popesso08,balestra10}, the VVDS survey \citep{lefevre04}, the
spectroscopic follow-up of the K20 survey
\citep{mignoli05,ravikumar07}. We reach a total of 238 spectroscopic
redshifts for the radio sources.

  We are able to extend our redshift list thanks to the deep and
  wide photometry available in this region of the sky.  We collect
  photometric redshifts from the catalogue based on photometric data
  from the far UV up to 8 $\mu$m (see Bonzini et al., in prep). We
  also use the MUSYC-E-CDFS catalogue \citep{cardamone10}, the
  GOODS-MUSIC catalogue \citep{santini09}, obtained using the excellent
  optical and NIR data in the GOODS region, and the COMBO-17 survey
  catalogue \citep{wolf04}. Combining spectroscopic and photometric
  redshifts for our sources, we end up with 1156 redshift for the
  X-ray sources ($\sim 85$\% of the total) and 641 for the radio
  sources ($\sim 73$\% of the total). Details are given in Table
  \ref{tab:tablez}.

\begin{table*}
 \centering
 \begin{minipage}{120mm}
 
 \begin{tabular}{ccccc}
 \multicolumn{5}{c}{Optically identified sources with redshift} \\
  Sample & \# of sources  & \# with spec z & \# with photo z & \# with z \\
  \hline
  \hline
  X-ray 4 Ms CDFS    & 732  & 418  & 661  & 666 (91\%) \\
  X-ray E-CDFS       & 753  & 414  & 609  & 627 (82\%) \\
  Unique X-ray       & 1352 & 706  & 1132 & 1156 (85\%)\\ 
  \hline
  Radio ($>$5$\sigma$) E-CDFS & 879 & 238 & 633 & 641 (73\%) \\ 
  \hline
 \end{tabular}
  \caption{Total available photometric and spectroscopic redshifts from the
  optical identification for both radio and X-ray samples.}\label{tab:tablez}

\end{minipage}
\end{table*}

\section{Finding X-ray counterparts of radio sources}\label{sec:match}

We combine the X-ray and radio information for our sources by
cross-correlating the radio catalog at $S/N > 4$ with the X-ray
catalogs of Xue et al. (2011) and \citet{lehmer05}. An X--ray source
is considered a candidate counterpart of a radio source if their
separation is less than 3$\sigma_d$, where $\sigma^2_d = \sigma^2_r +
\sigma^2_X$. The $\sigma_r$ and $\sigma_X$ are the \textit{rms} error
of the radio and X--ray positions. The radio rms is fixed at 0.2",
representative of the positional error across the field of view as we
discussed in Section \ref{sec:rdata}.  The X--ray rms ranges from 0.1"
to 1.5" for CDFS-4Ms sources and from 0.6" to 2.6" for E-CDFS sources,
according to Xue et al. (2011) and Lehmer et al. (2005), respectively.
In the case of multiple identifications, we simply choose the X-ray candidate
counterpart with a smaller separation.  This very simple matching
criterion is effective thanks to the relatively low surface density of
X-ray and radio sources, which implies a low rate of random
matches. Among the 1571 radio sources of the Miller catalog at $S/N >
4$, we identify 168 X-ray counterparts in the Xue et al. (2011)
catalog, and 175 in the Lehmer et al. (2005) catalog, with 66 sources
in common. To prevent the degradation of the data quality due to the
PSF at the edge of the CDFS-4 Ms, we consider the CDFS-4 Ms counterparts
only for off-axis angles smaller than 9.5 arcmin with respect to the
aimpoint.

We estimate the false matching rate due to random associations by
generating a set of radio catalogs with the same source density as that 
of Miller ($S/N > 4$), and match them with the X-ray catalogs in
CDFS and E-CDFS separately. We repeat this procedure 1000 times and
find an average random contamination of 3 spurious associations in the CDFS-4 Ms
%(~2\% of found matches)
and 8 in the E-CDFS area not covered by the CDFS-4 Ms survey. %(~7\%).

Therefore, we proceed with a refinement of the X-ray--radio cross correlation 
through the identification of the optical counterparts. 
The process of identifying the optical counterparts
must rely on a complex algorithm due to the larger source density
of the optical band, as shown in \citet{mainieri08}, 
and will be presented in Bonzini et al. (2011).  The optical identification has been already 
done by our group for the sources present in the 1Ms catalog and the 
\citet{kellermann08} radio catalog.  This allows us to check 
for mismatch in the optical counterpart.
For the new sources we perform a simple visual inspection
of the optical images (mostly observed by HST and WFI) for all the
candidate counterparts.  This procedure led us to remove 
2 candidates from the VLA--CDFS-4 Ms match-list and 6 candidates from the
VLA--E-CDFS. These 8 sources are flagged as false matches,
since the optical counterparts are clearly different for the radio and
X-ray sources. It is difficult to estimate the residual 
contamination due to random associations of sources with X-ray and radio detection,
however we argue that this is at most $\sim 3$\%, given that the number of false 
matches removed by visual inspection is close to the expected 
spurious matches estimated for our matching criterion.

To summarize, after a careful inspection of the X-ray, radio and
optical images we have:
\begin{itemize}
\item 
  268 radio sources with X-ray counterparts: 152
  from the CDFS catalog and 116 
  from the E--CDFS catalog (see Table \ref{tab:table2} and Table \ref{tab:table3});
\item 	
  693 radio sources from the $S/N > 5$ catalog without X-ray counterparts;
\item 
  1084  X-ray sources (in the  CDFS-4 Ms
  and in the E-CDFS catalogs) not associated with any radio source.
\end{itemize}

\begin{figure}
\centering
\includegraphics[scale=0.5]{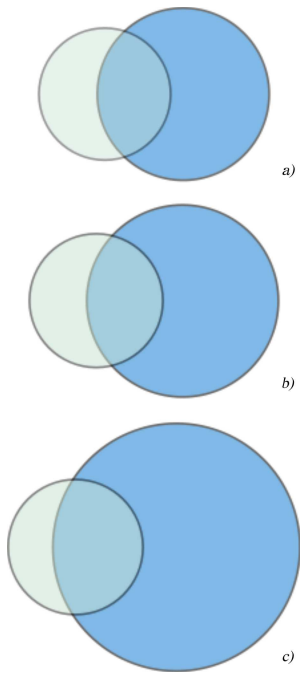}
\caption{Improvement in detected source number and overlapping of
  X-ray and radio sources in the central CDFS area. The X-ray
  detections (circles on the right) are respectively ({\it a}) from
  the 1Ms observation \citep{giacconi02}, ({\it b}) from the 2Ms
  observation \citep{luo08} and ({\it c}) from the 4Ms (Xue et
  al. 2011) observation. The radio sources from the 5 $\sigma$ catalog of
  Miller et al. (2011) are those falling in the central CDFS area
  (circles on the left).  The overlapping region represents the
  fraction of positional matches between X-ray and radio sources.}
\label{fig:m_diag}
\end{figure}

With the improvement in the depth of the X-ray and radio surveys we
can study how the number of X-ray counterparts of radio sources
changes with X-ray exposure in the CDFS area. In Figure
\ref{fig:m_diag} we compare three sets of observations of the CDFS in
the two bands with Venn diagrams. The first one is from the CDFS-1 Ms
X-ray observation (\citealt{giacconi02}, 366 sources) cross-matched
with the radio VLA observations from Miller et
al. (2011)\nocite{miller11} (S/N $>$ 5, 940 sources, of which 214 in
the CDFS).  The last two diagrams are the radio observations crossed
respectively with the 2 Ms (\citealt{luo08}, 462 sources) and 4 Ms
(Xue et al. 2011, 732 sources) observations.  Applying the positional
match with the CDFS-1 Ms and CDFS-2 Ms observations we adopt the same
algorithm used for the deeper 4 Ms and we find 105 and 112 matches
respectively. The number of radio sources with $S/N > 5$ with an
  X-ray counterpart in the CDFS-4 Ms observation is 131.  We note that
  the fraction of radio sources with X-ray counterpart increases from
  $\sim 50$\% in the 2 Ms up to $\sim 60$\% in the CDFS-4Ms
  observation.  If this trend is maintained, we expect to detect the
  X-ray counterparts of the large majority of the radio sources with a
  10-12 Ms exposure, even if are aware that the skycoverage does not
  increase uniformly with exposure time in all the CDFS area. This is confirmed
 by the fact that the stacked signal of all the radio-only sources in the CDFS
amount to an average of 9 photons per source in the soft band, therefore 
even an increase of a factor of 2-3 in the exposure time is sufficient to 
bring the large majority of the radio source above the X-ray detection threshold.

\section{Star forming galaxy candidates}

Our goal is to identify sources powered by star formation in the radio
and X-ray bands. We know that both radio and X-ray source populations
are dominated in number by AGN at high flux densities. However, in
both bands the fraction of SF galaxies increase and become dominant
toward low fluxes (see Paper III, Paper IV and Xue et al. 2011, Lehmer
et al. 2011, in preparation). Therefore, we expect to find a larger
number of SFG in our data set, with respect to previous studies.  We
adopt a conservative criterion by requiring each source to be
dominated by star formation processes both in the radio and the X-ray
band.  Our definition of SFG may not be in agreement with the
definition obtained in Paper V, where SFG are defined as those without
any signs of nuclear activity in any band.  Since we do not have the
IR information for all sources in our sample, we use only the radio,
X-ray and optical information when available.

\subsection{Methods of selection}

In order to be classified as a SFG, a source must satisfy several
criteria to guarantee that its emission is dominated by star formation 
in both bands.  Clearly, the X-ray spectral analysis we
provide in this Paper is more stringent than the radio diagnostics.
Therefore, we first apply the X-ray criteria, and then the radio
filter.

\begin{figure}
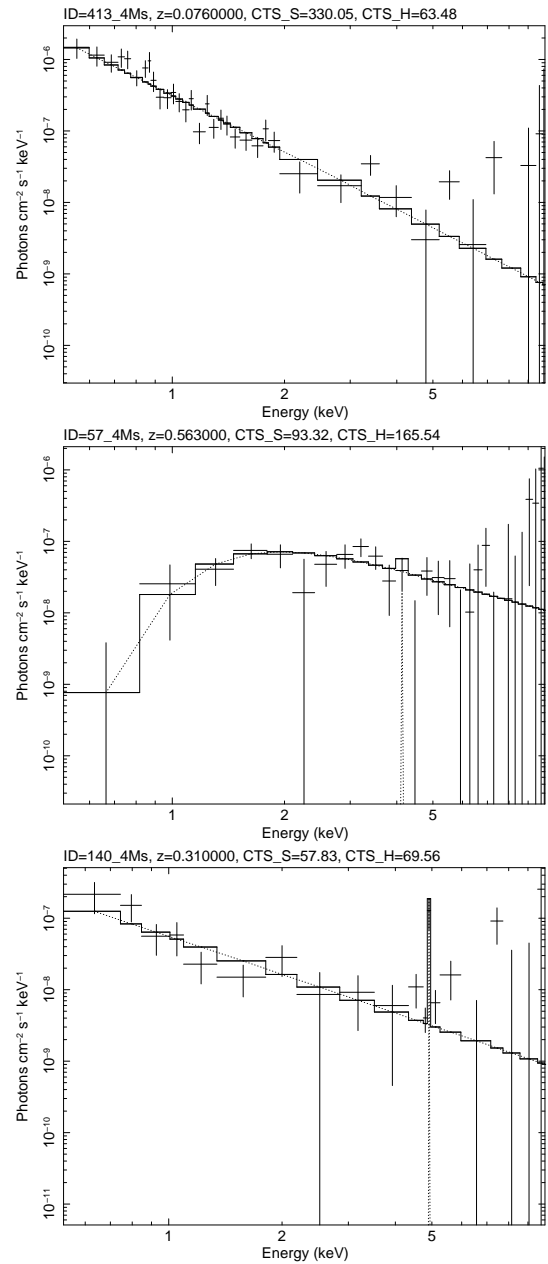

\centering
\includegraphics[scale=0.3,angle=270]{pluf_413_4Ms.ps} \\
\includegraphics[scale=0.3,angle=270]{pluf_57_4Ms.ps} \\
\includegraphics[scale=0.3,angle=270]{pluf_140_4Ms.ps} \\
\caption{ Examples of X-ray unfolded spectra for different type of
    sources. Top panel: a power law spectrum of a SF galaxy.  Middle
    panel: an absorbed AGN spectrum, where we show a clear intrinsic
    absorption at soft X-ray energies.  Bottom panel: an unabsorbed
    AGN spectrum with a redshifted Fe emission line. The source has
    been classified as AGN only because of the emission line, since it
    has low X-ray luminosity and absorption. }
\label{fig:fits}
\end{figure}

Traditionally, the main criterion to divide SFG from AGN is based on
X-ray luminosity, because of the limited range of emitted power of
star forming galaxies.  We adopt as a threshold $L_X = 10^{42}$ erg
s$^{-1}$ ($2-10 $ keV), where the unabsorbed X-ray luminosity is
obtained from the spectral fit. Clearly, the detection of the
  intrinsic absorption itself is already an important diagnostics.
SFGs are characterized by a power law X-ray spectrum (see top
  panel of Figure \ref{fig:fits}), without significant intrinsic
absorption \citep{norman04}.  Therefore, the detection of intrinsic
absorption with a significant confidence level (as in middle
  panel of Figure \ref{fig:fits} where we show an AGN spectrum clearly
  absorbed in the soft band) reveals that the X-ray flux is dominated
by nuclear emission (see \citealt{alexander05,brightman11}).  In
fact, when the X-ray emission is coming from regions distributed
across the galaxy, as in the case of star formation, only a negligible
amount of neutral gas screens the observed emission, as opposed to the
case of nuclear emission, where a limited mass of matter concentrated
towards the nucleus can result in a high value of $N_H$ (but see
  \citet{Iwasawa09} for a possible case of X-ray absorbed
  starburst). Therefore we consider as AGN candidates all the
sources with a column density higher than $N_H > 3 \times 10^{21}$
cm$^{-2}$ at least at one $\sigma$ confidence level (e.g. taking as
reference Figure 7 in \citealt{brightman11}).

Another strong indicator of nuclear activity is the presence of a
K-shell Fe line at 6.4 keV in the source spectra, as is often found in
Seyfert galaxies (e.g., \citealt{nandra94}).  We identify Fe lines by
adding an unresolved Gaussian component with energy $E=6.4/(1+z)$ keV,
where $z$ is fixed (see bottom panel of Figure
  \ref{fig:fits}).  The typical equivalent width of this emission
line is ~150 eV \citep{nandra94}. We cannot identify lines with
  such a low EW. Therefore, we decide to search only for the high tail
  of the distribution of line strength. In order to set a threshold 
  for the Fe line detection we run a set of spectral simulations 
  of powerlaw X-ray sources with S/N comparable to our sample. We find 
  that by considering only strong lines with an
  equivalent width larger than 1.5 keV, and sources with at
  least 10 counts in the hard band, we identify Fe line at 99\% of
  confidence level.

Another observational window useful to identify AGN is time
variability.  AGN are expected to show strong variability in their
X-ray light curves.  Since the CDFS 4 Ms data have been accumulated in
a time interval of about 11 years, we are able to search for X-ray
variability.  Clearly, star formation processes are not expected to
show timing variability, except at very low star formation rates,
where single compact sources may dominate the emission (see
\citealt{gilfanov04}).  Therefore, we will classify as AGN all the
sources with X-ray variability at a confidence level larger than 95\%
(see Paolillo et al. 2004, 2011\nocite{paolillo04,paolillo11}).

Another typical AGN indicator is the X-ray to optical flux ratio.
Using the optical fluxes from the WFI catalog available for
  the 75\% of X-ray CDFS sources and for 68\% of radio E-CDFS
  sources, we compute the F$_{X}$/F$_{opt}$ ratio
\footnote{Log(F$_{X}$/F$_{opt}$) is given by Log(F$_{0.5-2
    keV}$/F$_{opt}$) $\equiv$ Log(F$_{0.5-2 keV}$)+ 0.4$R$ + 5.71,
  where $R$ is the $R$-band magnitude \citep{szokoly04}.}.  Sources
with Log(F$_{X}$/F$_{opt}$) $> -1$ are assumed to be powered by
nuclear activity (see \citealt{bauer02,bauer04}). For the sources that
have no R-band detection, we use an R-band upper limit.

As for the radio band, we set an upper limit to the radio power
associated with star forming galaxies equal to $L_R = 5 \times
10^{23}$ W Hz$^{-1}$ (at 1.4 GHz).  We check that if we adopt the
local $L_X-L_R$ relation (Ranalli et al. 2003; Persic \& Rephaeli
2007) this limit roughly corresponds to the same star formation rate
level implied by the X-ray luminosity upper limit.

Thanks to two VLA observations we can explore also radio variability.
We identify sources with radio variability by comparing the radio
flux densities measured by \citet{kellermann08} with those from
Miller et al. (2011)\nocite{miller11} for the 261 common sources.  
We detect radio variability for 38 sources
whose difference between the two radio flux density
measurements is larger than three times
$\sigma=\sqrt{\sigma_{K}^2+\sigma_M^2}$, where $\sigma_{K}$ and $\sigma_M$
are the 1 sigma errors on flux densities in the catalogs of Kellermann 
and Miller, respectively.

We also have spectral information for a subset of 174 radio sources
with 5 GHz fluxes in the \citet{kellermann08} catalog.  The radio
slope is estimated $\alpha_R \equiv Log(F_{1.4GHz}/F_{5GHz})$. Normal
spiral galaxies are expected to have an average radio slope of 0.8,
while SN remnants and HII regions have a lower range of $\alpha_R$,
extending from 0.5 down to 0.1 \citep{gordon99}.  Then only very flat
spectrum radio sources, with $\alpha_R < -0.3$ (e.g., \citealt{li08}),
can be classified as AGN at high confidence level according to this
criterion.

Finally, we remove all the obvious radio sources which appear to be
the components of a multiple source in Miller et
al. (2011)\nocite{miller11}.  We also check visually the optical
images of all the sources which are seen as extended at a
3$\sigma$ confidence level.  This step allows us to identify sources
with Faranoff-Riley (FRI or FRII) morphology unless the diffuse radio
emission can be associated with the disc of relatively nearby
galaxies.

To summarize, our selection procedure consists in applying the
following criteria to the sources detected in both the radio and X-ray
bands:

\begin{itemize}
\item unabsorbed X-ray luminosity in hard band L$_{X}$(2-10 keV) $<
  10^{42}$ erg s$^{-1}$;
\item intrinsic absorption $N_H < 3 \times 10^{21}$ cm$^{-2}$;
\item no evidence of Fe emission line (EW$_{Fe} < 1.5$ keV) in sources
with more than 10 counts in the X-ray hard band;
\item no X-ray time variability in the sources of the CDFS-4Ms at 95\% confidence level;
\item X-ray to optical ratio Log(F$_{X}$/F$_{opt}$) $< -1$; 
\item radio power $L_{1.4 GHz} < 5 \times 10^{23}$ W Hz$^{-1}$;
\item no radio variability between the fluxes measured by \citet{kellermann08} and
Miller et al. (2011)\nocite{miller11} at a 3 $\sigma$ confidence level;
\item radio slope $\alpha_{R} > -0.3$, measured between 1.4 GHz and 5 GHz; 
\item no evidence in radio image of FR II or FR I morphology.
\end{itemize}

\subsection{Properties of radio sources with X--ray 
counterparts}\label{sec:XR}

\begin{figure}
\centering
\includegraphics[scale=0.44]{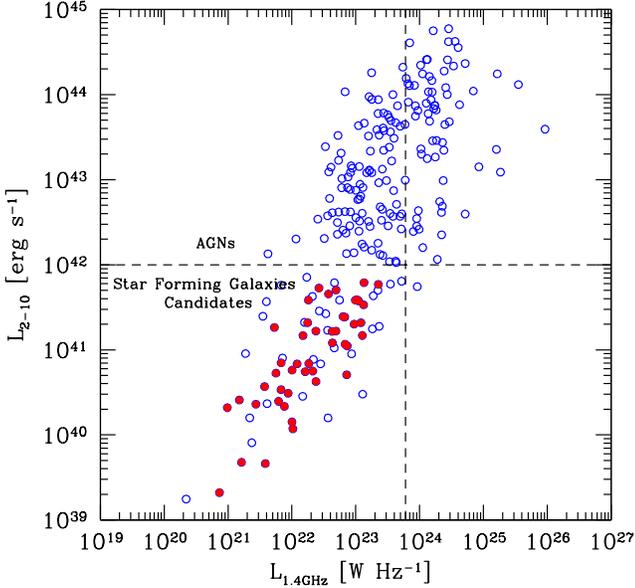}
\caption{Hard X--ray luminosity versus radio luminosity for all the
  257 sources detected both in X-ray and radio, and with
    measured redshift. The SF galaxy candidate sector is the bottom
  left identified with dashed lines.  Sources above the luminosity
  values indicated by the dashed lines are classified as AGN, while
  the lower left corner is populated with a mix of star forming
  galaxies and low luminosity AGN. Final SFG sources obtained with our
  full selection procedure are indicated with filled
  circles.}\label{fig:lh_lr}
\end{figure}

\begin{figure}
\centering
\includegraphics[scale=0.44]{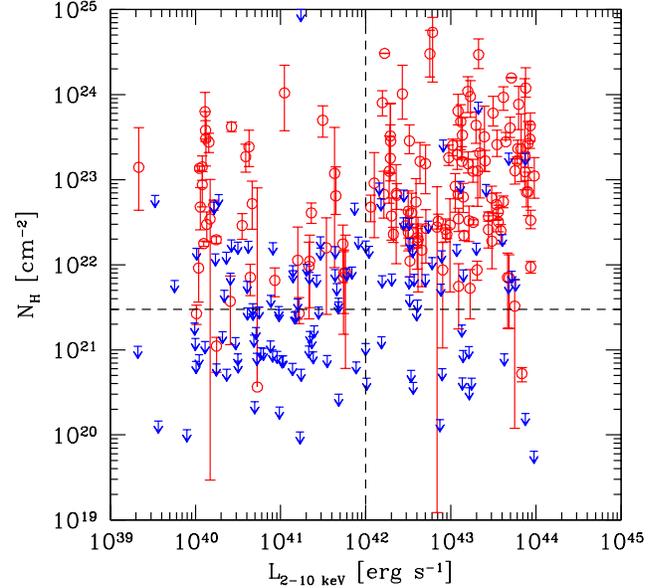}
\caption{Intrinsic absorption $N_{H}$ versus hard band intrinsic X-ray
  luminosity for all radio sources with an X-ray counterpart.  
  Error bars refer to 1 $\sigma$.  Arrows show 1 $\sigma$
  upper limits.  The horizontal and vertical dashed lines show the 
  thresholds in intrinsic absorption and hard X-ray luminosity between AGN and 
  SFG, respectively.
  }\label{fig:nh_lh}
\end{figure}

We perform the X-ray analysis based on the new 4 Ms data in the CDFS
for the sources within 9.5 arcmin of the aimpoint, while we use the
E-CDFS data otherwise. Among the 268 radio sources with an X-ray
counterpart, 257 ($\sim$ 95$\%$) have a spectroscopic or photometric
redshift. Most of the 11 sources without redshift lie on the
  edge of the optical coverage or near bright
  stars. Therefore, their redshift estimation is very difficult.  We
  decide to exclude these sources from the analysis. We check a
  posteriori that  including these sources assuming a redshift of
  1, which is the average redshift of our sample, does not affect
  significantly the analysis.  The intrinsic X-ray emission of the
sources is modeled with a power law with slope $\Gamma$, which is left
free only for spectra with more than 170 net counts, while it is
frozen to $\Gamma$=1.8 in cases of sources with fewer counts (as in
\citealt{tozzi06}). We model the intrinsic absorption with {\tt zwabs}
at the redshift of the source.  The local Galactic absorption is
modelled with {\tt tbabs}\footnote{For the {\tt zwabs} and the {tbabs}
  models see the {\sl Xspec} manual
  http://heasarc.nasa.gov/docs/xanadu/xspec/manual/manual.html} with a
column density fixed to $N_{H} = 8.9 \times 10^{19}$ cm$^{-2}$.  We
include the redshifted K--shell neutral Fe line modelled as a Gaussian
component with zero width at $6.4/(1 + z )$ keV.  We perform X-ray
spectral fits for almost all the sources in our sample, including
spectra with a net number of counts as low as 20. The rest--frame,
intrinsic X--ray luminosities are computed in the soft and hard band
from the best-fit models, after removing the intrinsic absorption.  In
some cases, the measured intrinsic absorption has a large uncertainty
which is accounted for by the statistical error estimated within {\sl
  Xspec}.  We take into account this uncertainty when measuring the
unabsorbed luminosity and associated error. The error on the X-ray
luminosity includes also the Poissonian uncertainty in the total net
detected counts.  Radio luminosities are computed as $L_{1.4GHz}=4\pi
d^2_LS_{1.4GHz}10^{-33}(1+z)^{\alpha_R-1}$ W Hz$^{-1}$, where $d_L$ is
the luminosity distance (cm) and $S_{1.4GHz}$ is the flux density
($mJy$), assuming an average value  $\alpha_R=0.7$ for the radio slope.

In Figure \ref{fig:lh_lr} we plot the intrinsic X--ray
2-10 kev (rest frame) luminosity versus the radio
luminosity for all 257 sources with X--ray and radio
detections with measured redshift. The lower left quadrant with 
$L_X < 10^{42}$ erg s$^{-1}$ and $L_R < 5 \times 10^{23}$ W Hz$^{-1}$
provides a first selection for the star 
forming galaxy candidates, based on the emitted power (76 sources).
We note that there is a large scatter along an apparent correlation 
between $L_X-L_R$ across five orders of magnitudes in luminosity. 
We also note that the upper limit on the X-ray luminosity is much 
more efficient in classifying star forming galaxies than the
threshold on the radio luminosity, since almost all sources 
screened by radio luminosity are already screened by the X-ray
luminosity criterion. 

In Figure \ref{fig:nh_lh} we plot the intrinsic absorption versus the
hard X-ray luminosity for all the radio sources with X-ray
counterpart.  Typically AGN can be classified in two groups on the
basis of their X-ray spectra, absorbed and unabsorbed AGNs, depending
whether their intrinsic absorption is above or below the threshold
$N_H = 10^{22}$ cm$^{-2}$.  In any case, as we discussed, a firm
detection of intrinsic absorption indicates that the X-ray emission is
powered by a nuclear source.  Therefore we flag as AGN all the sources
with $N_H > 3\times 10^{21}$ cm$^{-2}$ at least at one $\sigma$
confidence level, as shown in Figures \ref{fig:nh_lh}.  Then we check
the presence of the Fe K line, measuring equivalent width of the
Gaussian component at $6.4/(1 + z )$ keV.  A detailed X-ray spectral
analysis is important since it allows us to identify as AGN 24
  sources of those selected with luminosity below $10^{42}$ erg
s$^{-1}$, 13 because of the intrinsic absorption and
11 because of the Fe line feature. We also remove from our
SFG candidates sample all sources with evident variability in X-ray or
radio flux density, rejecting 3 sources in both cases.
Optical flux in the R-band is available for the 87\% of our
  sample.  The criterion on the X-ray to optical flux ratio, Log
  F$_X$/F$_{opt.} <$ -1, identifies 3 additional AGN.  The criterion
  on the radio slope $\alpha_R >$ -0.3 does not have any effect when
  applied after the other criteria.  As a further check we verify that
no sources have radio morphology typical of FR I. We show in
  Table \ref{tab:select} the effect of each single criterion, on the
  radio sources with X-ray counterpart.  We note that the most
effective criterion is the X-ray luminosity, and, overall, the use of
the selection based only on the X-ray data is strong enough to screen
$\sim 98$\% of AGN.  To summarize, we classify as star forming
galaxies 43 sources out of 257 on the basis of X-ray and
radio data.  The fraction of sources detected in both bands and
  classified as SFG ($\sim 17$\%) is much lower with respect to
  previous works based on shallower data in the same E--CDFS region
  (60\%, see \citealt{rovilos07a}). This is 
  %mostly 
  due to our
  more conservative selection criteria, despite that we do not include 
  IR data.

\begin{table}
 \centering
 \begin{minipage}{80mm}
  \centering
   \caption{SFG and AGN are the number of galaxies identified as
     powered by star formation or nuclear activity, respectively, in
     the sample of 257 sources detected in both bands,
     according to single selection criteria.  }
  \begin{minipage}{65mm}
  \centering
 \begin{tabular}{lccc}
  \hline
Selection criterion & SFG & AGN \\%& Efficiency \\
  \hline
  \hline
 X-ray luminosity & 78  & 179 \\%& \textbf{83\% }\\
 Column density   & 131 & 126 \\%& \textbf{54\% }\\
 Fe line          & 229 & 28  \\%& \textbf{8\% }\\
 X-ray variability & 212 & 45 \\%&\textbf{ 16\% }\\
 F$_{X}$/F$_{opt}$ & 148 & 109 \\%& \textbf{46\% }\\
 \hline
 All X-ray criteria & 47 & 210 \\%&\textbf{ 98\% }\\
 \hline
 Radio luminosity & 172 & 85 \\%& \textbf{40\% }\\
% Radio slope $\alpha_R$ & & & \\
 Radio variability\footnote{The numbers refer only to the sample with this information available.} & 99 & 15 \\%& \textbf{13\% }\\
  \hline
 All radio criteria & 164 & 93 \\%& \textbf{43\% }\\ 
  \hline
 \end{tabular}
 \end{minipage}
\label{tab:select}
\end{minipage}
\end{table}

We compare our classification method with that of \citet{padovani11}
for the 73 sources we have in common.  In Papers V, the classification
method relies also on optical and IR photometry, and, when available,
optical spectroscopy and morphology.  Among the sample of star forming
galaxy candidates, 6 sources are identified as objects powered by
nuclear activity by \citet{padovani11}.  These 6 sources are
classified as AGN by criteria or bands not available for all our
sample, where additional signatures of nuclear activity may be
observed.  Specifically, all of them have a $q$ ratio\footnote{The
  value of $q$ is calculated as the ratio of far-IR to 1.4GHz flux
  density $q=log{[(FIR/3.75 \times 10^{12})/S_{1.4GHz}]}$.} lower than
1.7, the nominal dividing value between SFGs ($q>1.7$) and AGNs
($q<1.7$). We decide to keep these sources in our SF candidate
  sample, since both X-ray and radio data are still consistent with
  being powered by star formation only.

\subsection{Star forming galaxy candidates with X-ray
detection only}\label{sec:XO}

\begin{figure}
\centering
\includegraphics[scale=0.44]{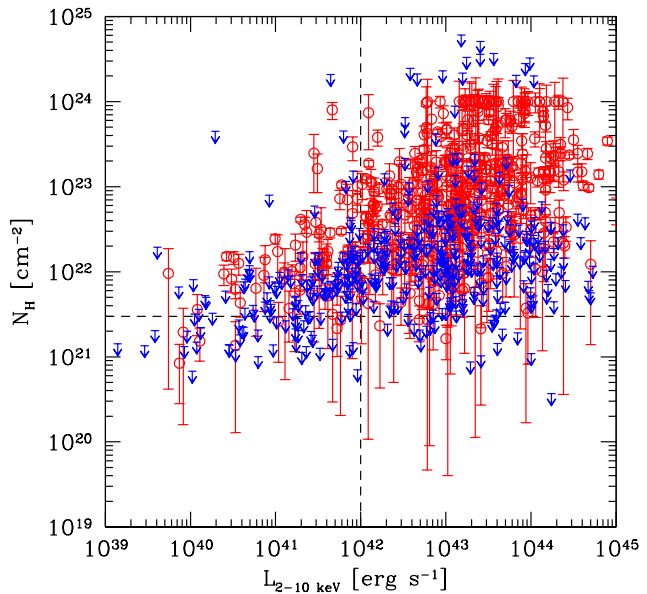}
\caption{Intrinsic absorption $N_{H}$ versus hard band X-ray
  luminosity in the sample of X-ray sources with no detection in the
  radio.  Error bars refer to 1 $\sigma$.  The horizontal dashed line
  shows the threshold between AGN and SFG, $N_H = 3 \times 10^{21}$
  cm$^{-2}$, while the vertical one shows the nominal dividing X-ray
  power between star forming galaxies and AGN.}\label{fig:nh_lh_x}
\end{figure}

Here we check for SFG among the X-ray sources without radio
detections.  Clearly, only the X-ray criteria can be applied. In the 4
Ms and E--CDFS catalogs there are respectively 564 and 587 sources
detected only in the X--ray band. Excluding common detections in the
two catalogs in the central part of the field and 10 sources already
classified as stars, there are 1074 X--ray extragalactic sources with
no radio counterpart in the E--CDFS area. Among them, 899 ($\sim 83$
\%) have optical spectroscopic or photometric redshifts.  For these
sources we can apply the same criteria based on $L_X$ and the
properties extracted from the spectral analysis, $N_H$, the equivalent
width of the Fe line and F$_X$/F$_{opt}$.
Given the
negligible effect of the upper limit on the radio luminosity used in
Section 4.1, we notice that these sources are selected with
essentially the same criteria as those with a radio counterpart except
for radio variability.

The intrinsic absorption of the sample of sources detected only in the 
X-ray is shown in Figure \ref{fig:nh_lh_x}. In the subsample of 
899 sources with known redshift, 208 ($\sim$23$\%$) have an X--ray
luminosity below $10^{42}$ erg s$^{-1}$. It is also evident that lager
values of intrinsic absorption becomes common at high luminosity. Applying 
the criteria based on X-ray hard luminosity, intrinsic absorption, Fe K 
line presence, variability and the F$_X$/F$_{opt.}$ ratio, using the R-band 
magnitudes from \citet{xue11} and \citet{lehmer05},
we identify 70 SF galaxy candidates. 
Therefore we add 70 sources classified as SF galaxy candidates,
with upper limits in the radio band obtained by measuring flux density
 in the VLA image at the position of the X-ray source (see Miller et al. 
2011\nocite{miller11}). For these sources we use $3 \sigma$ radio upper 
limits at the X-ray position, and we will
take them into account in the censored data analysis of the $L_R$-$L_X$
correlation in section \ref{sec:lrlx}.

\subsection{Star forming galaxy candidate with radio
detection only}\label{sec:RO}

In this section we analyze all the radio sources with no X-ray
detections in the E-CDFS. In our radio catalog (S/N $>$ 5) we find
705 single sources, excluding multiple components, with
  radio-only detections.  Among them 423 ($\sim$ 60 \%) have
  spectroscopic or photometric redshifts.  We evaluate the X-ray
upper limits of these sources by performing photometry on the X-ray
image after removing all the X-ray detected sources.  Then we
calculate the conversion factor between the vignetting-corrected count
rate and the flux, assuming a simple power law spectrum without
intrinsic absorption, and evaluate the luminosity at each radio source
position.  Due to the impossibility of performing a complete spectral
analysis we cannot rely on additional X-ray indicators of nuclear
activity such as strong absorption and Fe emission lines. We select as
star forming galaxy candidates 117 objects with radio
luminosity lower than $5 \times 10^{23}$ W Hz$^{-1}$ and 3 $\sigma$
upper limit on X-ray luminosity lower than $10^{42}$ erg s$^{-1}$.  We
removed 4 sources showing radio variability and 2 because of
  the presence of jets. We also checked for consistency with the
multiwavelength classification obtained in Paper V for a subsample of
these sources.  We find a total of 9 SF galaxies in the radio
only sample that were classified as AGNs in Paper V. Again we decide
to keep these sources.  As a further check we stack X-ray images
  of radio sources in three redshift bins. The stacked signal is null
  in the hard band, therefore the average hardness ratio
  (HR)\footnote{The hardness ratio is defined as HR=(H-S)/(H+S), where
    S and H are the soft and hard X-ray net counts corrected for
    vignetting.} of the radio sources is non measurable but still
  consistent with a soft spectrum. Since the selection of this
  sample is weaker than the previous ones, due to the lack of X-ray
  detections, a further screening should be done with a
  multiwavelength analysis. Nevertheless, we select a total of 111 SFG
  candidates with redshifts among the radio only sample, and we will
  consider them in the correlation analysis.

\section{The $L_R$-$L_X$ correlation for SF galaxies at high redshift}\label{sec:lrlx}

\begin{figure}
\centering
\includegraphics[scale=0.44]{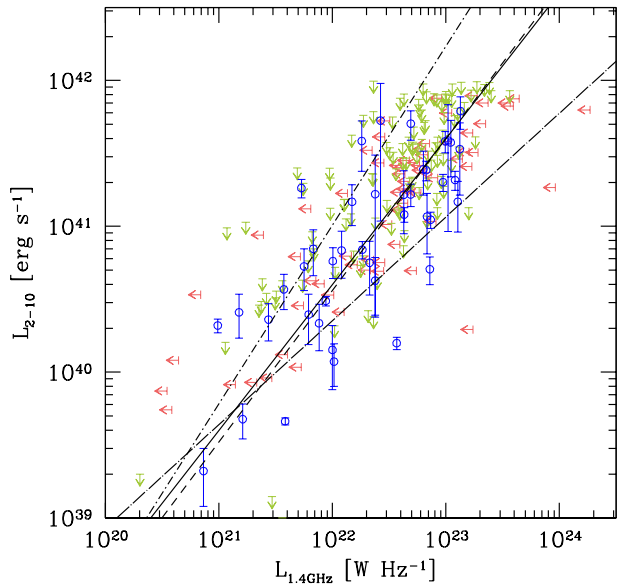}
\caption{Radio X-ray luminosity relation for all the star forming
  candidates. The open circles are the radio sources with X-ray
  counterparts. Error bars refer to 1 $\sigma$. The arrows are the
  radio and X-ray 1 $\sigma$ upper limits, respectively.  The four
  lines shows Eqs. 1-4. The solid line is the best-fit of the data
  detected in both bands assuming a slope of unity. The short dashed
  line is the linear fit of matches without a fixed slope. The
    dot-short-dashed and dot-long dashed lines are the best fit
    obtained with censored analysis using also the sources detected
    only in the X-ray or radio band, respectively, with upper limits
    in the other band.}\label{fig:lr_lh_asurv}
\end{figure}

\begin{figure}
\centering
\includegraphics[scale=0.44]{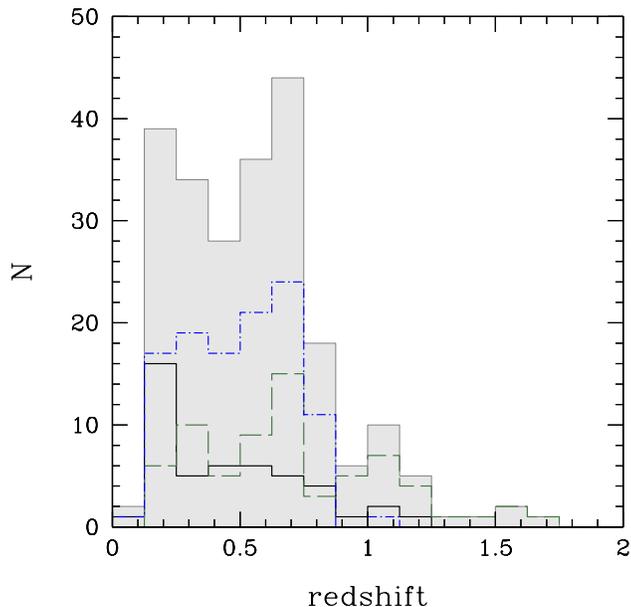}
\caption{Redshift distribution of star forming galaxy candidates
  selected in the three samples of sources. The solid line is the
  histogram of sources detected in both X-ray and radio bands. The
  dashed line is the histogram of sources detected only in the X-ray
  band and the dot-dashed line is the histogram of sources detected
  only in the radio band.  The shaded histogram is the distribution of
  the total sample.}\label{fig:SFG_z_hist}
\end{figure}

  We selected SF galaxy candidates by requiring that both X-ray and
  radio properties are consistent with being powered by star
  formation.  This is clearly a conservative estimate, since the
  presence of nuclear emission in one band implies the rejection of
  the sources, even if, in principle, the emission in the other band
  may not be contaminated by the nuclear emission. For example,
    in Paper V, radio quiet AGN have been found consistent with being
    powered by star formation in the radio band.  We do not aim at
    obtaining a complete census of star formation, therefore we remark
    that our selection does not allow us to trace directly the cosmic
    star formation rate.

  To investigate the $L_R$-$L_X$ relation at high redshift we
  use the sample of 43 star forming galaxies with detections
  in both bands, together with the 70 sources detected in the
  X-ray band only and the 111 detected only in the radio.  We
  find a strong correlation (the coefficient of Spearman test is 0.75, with a
  null hypothesis probability of 6$\times$10$^{-9}$) over almost three
  orders of magnitude for sources detected in both bands.

  Assuming a linear relation with the slope fixed to unity, similar to
  what was done by \citet{ranalli03}, we find
  \begin{small}
  \begin{equation}
  Log(L_{2-10keV})=Log(L_{1.4GHz})+18.60 \pm 0.44, 
  \end{equation}
  \end{small}
  where $L_{2-10keV}$ is in erg s$^{-1}$ and $L_{1.4GHz}$ is in W
    Hz$^{-1}$ (see Figure \ref{fig:lr_lh_asurv}, solid line).

  If we leave the slope free to vary, we find:
  \begin{small}
  \begin{equation}
  Log(L_{2-10keV}) = (1.04 \pm 0.05)\times Log(L_{1.4GHz}) + (17.68 \pm 1.15) \, 
  \end{equation}
  \end{small}
  obtained by performing two least-squares regressions assuming
  alternatively $L_{2-10keV}$ or $L_{1.4GHz}$ as independent variable,
  and then using the bisector method \citep{isobe90}. This
    relation is consistent with the fit found fixing the slope to
    unity (see Figure \ref{fig:lr_lh_asurv}, dashed line). Moreover,
    it is consistent with the linear relation found for local SF
    galaxies.  We are aware that the relation may not hold for low
  luminosity galaxies, where the SFR may not be proportional to the
  HMXB population \citep{grimm03}.

  We note a large uncertainty in the normalization of this relation
  (a factor from 3 to 10 assuming a fixed slope or a free slope,
  respectively, at 1 $\sigma$ level).  This uncertainty is largely due
  to the significant scatter in the observed relation which appears to
  be larger than that observed for local SF galaxies. The standard
  deviation of the X-ray luminosity with respect to the average
  relation is 0.41 dex, as opposed to 0.24 dex found in
  \citet{ranalli03}.  Since the average 1 $\sigma$ error of our X-ray
  luminosities is 0.06 dex, we conclude that the scatter is mostly due
  to an intrinsic component. The scatter can be due to an increasing
  X-ray emission component proportional to stellar mass and not to the
  instantaneous SFR, analogous to what has been found in a sample
  of luminous infrared galaxies by \citet{lehmer10}.

  We now consider the star forming candidates detected only in one
  band, including the upper limits at 3 sigma in the other band.  We
  use the software for statistical analysis ASURV
  \citep{feigelson85,isobe86}.  This tool allows us to perform
  censored data analysis, including first upper limits in the radio
  band and then separately the upper limits in X-ray band.  The
  best-fits for the $L_R$-$L_X$ relation, adding X-ray--only
  sources or radio-only sources, are, respectively:
  \begin{small}
  \begin{equation}
  Log(L_{2-10keV}) = (1.22 \pm 0.15)\times Log(L_{1.4GHz}) + (14.02
  \pm 3.22) \, 
  \end{equation}
  \end{small}
  \begin{small}
  \begin{equation}
  Log(L_{2-10keV}) = (0.71 \pm 0.08)\times Log(L_{1.4GHz}) + (24.70
  \pm 1.74) \, .
  \end{equation}
  \end{small}
    In both cases an X-ray/radio correlation is detected at more
    than 3 sigma confidence level.  The fit which includes X-ray--only
    sources is consistent with a slope equal to 1.  On the contrary,
    the slope found including radio--only sources is inconsistent with
    unity.  This result is similar to that found by \citet{lehmer10}
    for SFR $\gtrsim 0.4 M_{\odot} yr^{-1}$, suggesting that the
    dependence of $L_X$ on SFR is not linear.  Deeper X-ray data and
    radio would help to obtain more robust results for our
    high-redshift galaxies.

  For star forming candidates detected in both bands, we also
  check if there is some evolution with redshift in the relation, by
  separating low redshift ($z < 0.5$) objects from high redshift ($z
  \geq 0.5$ ) ones. We find that the intercept varies from
    18.66 for low redshift sources to 18.47 for high redshift sources,
    with an uncertainty of 0.4 dex in both cases, therefore showing no
    evolution. This confirms the claim of \citet{bauer02} based on a
    smaller sample of SF galaxies in the CDFN.  In Figure
  \ref{fig:SFG_z_hist} we show the redshift distribution of our star
  forming galaxy candidates. It shows that most of SF galaxies are
  distributed up to redshift 1, with a small fraction ($\sim$15$\%$)
  at higher redshift. The average redshift is $<z> \sim 0.5$, while
  $<z>_{matches} \sim 0.4$ for the sources detected in both
  bands. While AGNs are distributed up to high redshifts, star forming
  galaxies can be observed in the radio and X-ray with current data
  only up to redshift 1.5-2, as discussed in Section \ref{sec:exp}.

   To summarize, we do find a clear correlation between the X-ray
    and radio luminosity for star forming galaxies.  This result is
    robust when we include upper limits in the X-ray or radio bands in
    a censored data analysis, at variance with the claim by
    \citet{barger07}.  However, the observed correlation at high
    redshift appears to have a large intrinsic scatter, and its slope
    is still unclear. Both these features may be due to an increasing
    X-ray contribution proportional to the total stellar mass and
    unrelated to the instantaneous SFR.  In the following sections, we
    make use of the best-fit linear relation (Eq. 1) to estimate the
    star formation rate of our sources.

\section{Star Formation Rate}\label{sec:SFR}

To derive the average $L_X$-SFR relation in the selected sample of
star-forming galaxies, we use the measured $L_X$-$L_R$ relation and the
local-calibrated SFR-$L_R$.  We adopt the calibration between $L_R$
and SFR found by \citet{bell03}:

\begin{small}
\begin{equation}
SFR=5.52 \times 10^{-22} L_{1.4 GHz} M_{\odot}/yr \, ,
\label{eq:bell}
\end{equation}
\end{small}
where $L_{1.4 GHz}$ is in W Hz$^{-1}$.  Assuming Eq. (1) and
substituting into Eq. (5) we find:

\begin{small}
\begin{equation}
SFR = (1.40 \pm 0.32) \times  \frac{L_{2-10keV}} {10^{40} } M_{\odot}/yr \, ,
\label{eq:SFRX}
\end{equation}
\end{small}

\noindent
where $L_{2-10keV}$ is in erg s$^{-1}$. This relation is a factor of 2
lower than that obtained by \citet{persic07} for their low redshift
sample, where they found a factor 2.6$\times 10^{-40}$ between the SFR
and the X-ray hard luminosity, but still consistent within the
  uncertainty. We list our estimated SFR in Table \ref{tab:table1}
  along with the X-ray and radio luminosity.

\begin{figure}
\centering
\includegraphics[scale=0.44]{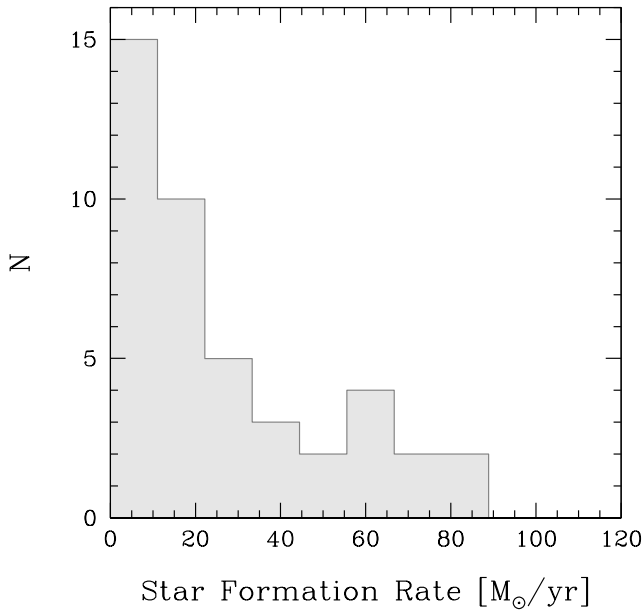}
\caption{Star formation rate distribution SF galaxy candidates with
  X-ray and radio detection.  The SFR is evaluated using the SFR-L$_R$
  relation.  }\label{fig:sfr_hist}
\end{figure}

  In Figure \ref{fig:sfr_hist} we show the distribution of the
  estimated SF rates in our sample derived directly from their radio
  luminosity. Most of SFGs (60-70 $\%$) have star formation rate lower
  than 30 $M_\odot yr^{-1}$.  About 20\% of galaxies have SFR which
  exceeds $50 M_\odot yr^{-1}$.  This implies a mix of different type
  of SFG, from normal spirals to strong starburst, as we discuss in
  the next section.

\subsection{Comparison with models}\label{sec:mod}

\begin{figure}
\centering
\includegraphics[scale=0.44]{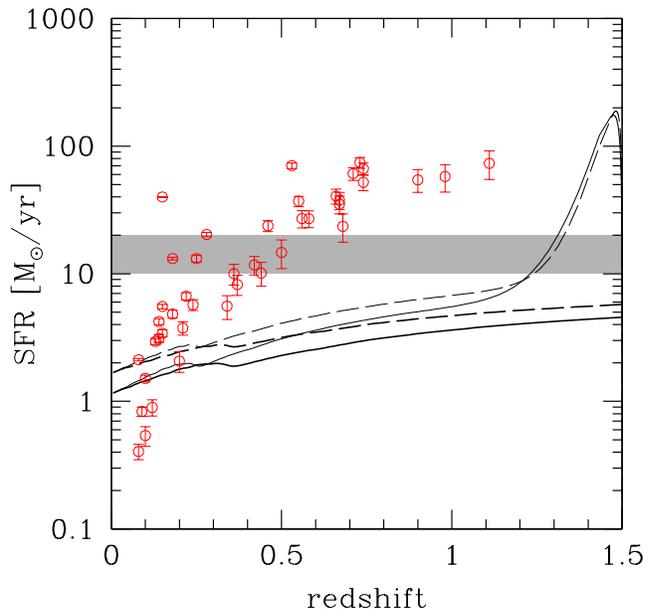}
\caption{SFR evaluated in our sample of galaxies detected in both
  X-ray and radio. The lines represent the different models compared
  to the data. The solid and dashed thick lines represent respectively
  the SFR in the Milky Way and M31, if we assume that the star
  formation started at z$_f=10$. The thin curves are the same
  galaxies in the case of a formation time of
  z$_f$=1.5. The horizontal stripe represents the observed SFR
  in the typical local starburst galaxy M82.}\label{fig:sfr_mw}
\end{figure}

We use chemical evolution models to predict the typical star formation
rate as a function of redshift in different galaxy types.  In
particular, we took into account the predictions from a model devised
for the Milky Way \citep{cescutti07}. This model is based on the
two--infall model of \citet{chiappini97}, which assumes that the Milky
Way formed in two major steps by means of gas accretion. The first gas
accretion episode formed the halo and the thick disc on a timescale
not exceeding 2 Gyr, whereas the second episode formed the thin disc
on a much longer timescale (e.g., 8 Gyr at the solar ring).  This
model accurately reproduces many properties of the Galactic disc and
therefore it can be used as an example of a normal spiral.  The star
formation rate is obtained by assuming a dependence on the surface gas
density $\Sigma_{gas}$ as suggested by \citet{kennicutt98}; in
particular, a dependence $ SFR \propto \Sigma_{gas}^k$ with $k=1.5$
and a threshold $\Sigma_{gas} = 7M_{\odot}pc^{-2}$ in the gas density
are assumed.  The total star formation rate is the sum of that
  occurring in the bulge and that occurring in the thin disc.  The
  model for the bulge of the Milky Way assumes that the SFR was much
  more intense and faster than in the discs.  The model also accounts
  for a varying SFR decreasing with galactocentric distance.  We
assume that the star formation started at redshift $z_f=10$.  A similar
model has been computed for the bulge and the disc of M31 which has a
mass roughly two times higher than the Milky Way.  These models both
reproduce the main properties of the bulges of the Milky Way and M31
(Ballero et al. 2007a, 2007b) \nocite{ballero07a,ballero07b}.  In
Figure \ref{fig:sfr_mw} we show the total predicted SFR (thick lines)
compared with our measured star formation rates as a function of
redshift. We note that for $z<1.5$ only the disc component is
  visible since the star formation in the bulge ended at high
  redshift.  The disc SFR for the Milky Way and M31 reaches values
comparable with the lowest found in our analysis, showing that the
majority of our star forming galaxies form stars ten times more
efficiently than Milky Way-like objects, up to a level comparable to
the strong starburst M82 (horizontal stripe). Obviously this is
  due to the flux limit of our survey which allows us to detect Milky
  Way-like only for $z<0.3$. We also note that a significant
  fraction of our galaxies reach values higher that that of M82 and
  comparable with the model predictions for formation of the Galactic
  bulge.  This is shown in Figure \ref{fig:sfr_mw} by the thin lines,
  which have been obtained by setting a formation time of $z_f=$1.5.
Although this is not realistic, since in true bulges are generally
old, it suggests that our star forming galaxies span a wide range of
starburst, from normal discs to strong burst typical of bulge
formation.

\subsection{Prospects for star forming galaxy survey}\label{sec:exp}

\begin{figure}
\centering
\includegraphics[scale=0.44]{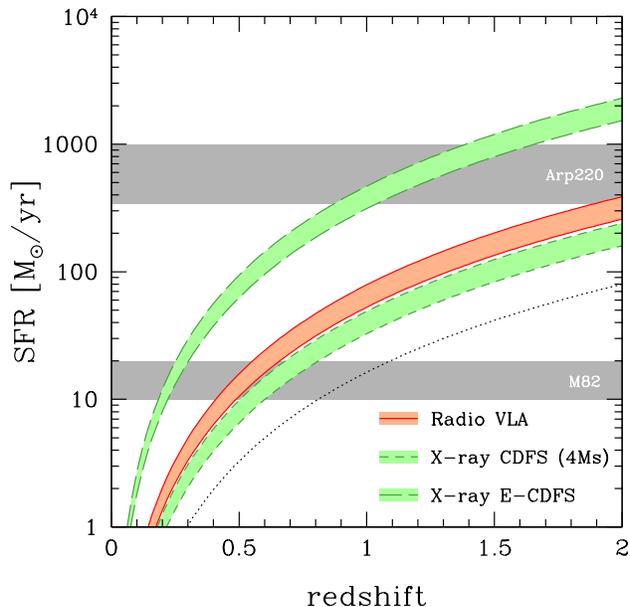}
\caption{Minimum measurable star formation rate as a function of
  redshift in the X-ray CDFS 4Ms, E-CDFS and radio surveys.  The SFR
  value corresponds to the hard band flux limit and to the 1.4 GHz
  flux density limit in the X-ray and radio band respectively,
  therefore it cannot be applied to the entire E-CDFS field.  We
  assume the \citet{bell03} correlation and Eq. 6 between the SFR and
  the radio and X-ray luminosity respectively, with an average
  uncertainty of 10\%. The horizontal stripes represent the observed
  SFR in the typical local starburst galaxies M82 and Arp220. The dotted line
    shows the limit that a 10Ms Chandra exposure in the CDFS 
    would be able to reach.}\label{fig:sfr_l}
\end{figure}

Due to the tight relation observed for the radio and X-ray emission of
star forming galaxies, we expect to see all the sources powered by
star formation in both bands, when an adequate sensitivity limit is
reached.  Assuming the linear correlations between the SFR and the
radio and X-ray luminosity, as evaluated by \citet{bell03} and our
Eq. 6, respectively, we compute a minimum detectable SFR at any given
redshift. Taking the lowest limit in the X-ray and radio CDFS surveys,
we plot in Figure \ref{fig:sfr_l} the minimum star formation rate that
we can detect in these two surveys.  We assumed a 10\% accuracy in
both cases in the relation between luminosity and star formation rate.

If we refer to M82 as a typical starburst galaxy in the local
universe, we immediately see the redshift range that we are actually
exploring.  M82 has a SFR of $10-20 M_\odot yr^{-1}$, corresponding to
more than 10 times that in the Milky Way at the present time (e.g.,
\citealt{doane93}).  A similar starburst galaxy can be seen up to $z
\sim 0.2$ in the X-ray E-CDFS, and up to $z \sim 0.6$ with the radio
VLA and in the 4 Ms data of the CDFS (note that this holds only for
the part of the field which reaches the most sensitive levels).  As a
consequence, lower values, typical of normal galaxies ($1-10 M_\odot
yr^{-1}$) can be seen only at moderate redshift $z<0.4$.  In the
strong starburst regime ($10-100 M_\odot yr^{-1}$) it is possible to
reach redshift as high as $z\sim 1.5$.  At larger redshift, only the
bright end of the SFG population is observable, i.e. starburst
galaxies with extremely large SF rates $>100 M_\odot yr^{-1}$. A
typical example of the latter population is Arp220, a nearby system in
final stages of galaxy merger with a powerful burst of star formation
at each of the nuclei and close to its peak in activity, which results
in strong radio, FIR and X-ray emission. Models of the embedded
starburst suggest a SFR$= 340-10^3$ $M_\odot yr^{-1}$, depending on
the assumed star formation tracer \citep{baan07}, observable in our
surveys up to $z \gtrsim 2$. To explore the range $> 100 M_\odot
  yr^{-1}$ up to $z=2$ is necessary to reach a factor $\sim 2.5$
  higher sensitivity (corresponding to the ideal 10 Ms goal for the
  CDFS in the X-ray band).  At the same time we would reach a complete
  census of strong starbursts (from M82 on) up to redshift 1.

\section{Conclusions}\label{sec:end}

Both radio and X-ray emission can be a clue to ongoing star formation
activity with the advantage of not being absorbed by dust.  In
particular, the collective emission in the radio band is produced by H
II regions and supernovae, while in the X-ray band originates from the
HMXBs and $\sim 1 $ keV gas heated by stellar winds.  In the
assumption that both bands always act as a calorimeter for the
emission from massive, short-lived stars, we expect to observe a well
defined correlation between X-ray and radio luminosity up to high
redshift.

In this paper we aim at exploring such a correlation using the deep
X-ray and radio data in the E--CDFS area (Lehmer et al. 2005; Xue et 
al. 2011; Miller et al. 2011)
\nocite{lehmer05,xue11,miller11}.  The starting sample includes $\sim
1300$ X-ray sources and $\sim 900$ radio sources.  Many of them have
an identified optical or IR counterpart (Xue et al. 2011; Bonzini et
al. in preparation).  A spectroscopic or photometric redshift is known
for 85\% and 73\% of the X-ray and radio sources, respectively.
First, we identify all the sources which are detected in both bands.
Then, on the basis of the radio data and of a detailed X-ray
spectroscopy we select a sample of star forming galaxy candidates.
This allows us to investigate the correlation between the radio and
X-ray luminosity, and to extend the measure of the star formation rate
for our sources over a large redshift range up to 1.5. Our results are
summarized as follows:

\begin{itemize}
 \item Among the 268 sources with both radio and X--ray detections,
   most of which have spectroscopic or photometric redshifts, 43
   ($\sim$16\%) are consistent with being powered by star formation
   processes in both bands.  The main criteria are the X--ray and radio
   luminosity, the intrinsic absorption, the presence of the Fe
   emission line, and the time variability. Among the sources
   detected only in the X-ray or the radio band, we select 70 and 111
   star forming candidates, respectively.  Their classification,
   though, being based on a smaller number of criteria, is more
   uncertain;

 \item We find that for the sources detected in both bands the
   $L_X-L_R$ relation is well fitted with a slope close to 1:
   $Log(L_{2-10keV}) = (1.04 \pm 0.05)\times Log(L_{1.4GHz}) + (17.68
   \pm 1.15)$. The fit which includes X-ray upper limits, treated
     with censored data analysis, is consistent with the previous
     relation. On the contrary, the inclusion of upper limits in the
     radio band leads to a flatter relation, which is suggesting of a
     non linear dependence of $L_X$ on instantaneous star
     formation. Deeper data are needed to obtain more robust results
     at high redshift;

\item Assuming a linear slope in the $L_X-L_R$ relation and splitting
  our sample in low and high redshift bins, we find no evolution in redshift;
  
\item We find that the $L_X-L_R$ relation shows a significant
  scatter.  We estimate its intrinsic component to be 0.4 dex,
  possibly due to a contribution to X-ray luminosity unrelated to the
  instantaneous SFR;

 \item Finally we compute relation between SFR and X-ray luminosity in
   the 2-10 keV band: $SFR = (1.40 \pm 0.32) \times L_{2-10keV}$
   $10^{-40}$ M$_{\odot}$/yr.  Most of the sources (~60$\%$) have a
   SFR lower than 30 M$_{\odot}$/yr. A small number sources have a SFR
   higher than 50-60 M$_{\odot}$/yr. The comparison of these data
     with models of chemical evolution allows us to explore the nature
     of the SFG galaxies. The SFRs we measure in our deep narrow
   survey span a wide range from normal spirals like the Milky Way or
   M31, to starburst like M82, up to strong starburst typical of
   bulges and spheroids in formation.
\end{itemize}

  The strong correlation of SFR with the hard X-ray luminosity in
  our high-z galaxy sample shows that X-ray surveys can provide a
  powerful and independent tool in measuring the instantaneous SFR in
  distant galaxies.  However, our data also indicates that the complex
  physics behind the X-ray and radio emission associated to star
  formation, may introduce significant scatter between $L_X$ and SFR.

\section*{Acknowledgments}
SV and PT acknowledge support under the contract ASI/INAF I/009/10/0.
WNB and YQX acknowledge support under the Chandra X-ray Center grant SP1-12007A and
NASA ADAP grant NNX10AC99G. We thank Ginevra Trinchieri and Anna Wolter
for useful discussion on the X-ray emission from star forming galaxies.
We also thanks Marcella Brusa for discussions.
The VLA is a facility of the National Radio Astronomy Observatory which is 
operated by Associated Universities Inc., under a cooperative agreement with 
the National Science Foundation.
\bibliographystyle{mn2e}
\bibliography{refs}

\begin{table*}
 \centering
 \begin{minipage}{120mm}
  \caption{X--ray spectral analysis results of sources with radio
  counterparts in the CDFS area. 
  X ID are from Xue et al. (2011);  
  z is the spectroscopic or photometric redshift;  
  $\Gamma$ is the slope of the intrinsic
  X-ray emission modelled with a power law; $N_H$ is the column density; $L_{0.5-2keV}$
  and $L_{2-10keV}$ are the luminosities in the soft and X-ray band
  respectively.\label{tab:table2}}
  \begin{tabular}{@{}rrrrrrr@{}}
  \hline
  \hline
  Xue X ID & z & $\Gamma$ & $N_H$ ($10^{22}$ cm$^{-2}$) & $L_{0.5-2keV}$ erg s$^{-1}$ &  $L_{2-10keV}$ erg s$^{-1}$ \\
  \hline
16 & 1.94 & 2.1 $ \pm $ 0.4 & $<$ 2.27  & $ 1.03 \pm 0.19 \times 10^{43} $ & $ 1.76 \pm 0.32 \times 10^{43} $ \\ 
18 & 2.14 & 1.2 $ \pm $ 0.7 &  $ 64 ^{+ 37 }_{ -20 } $ & $ 1.04 \pm 0.19 \times 10^{43} $ & $ 5.88 \pm 1.09 \times 10^{43} $ \\ 
23 & 0.18 & 1.8  & $<$ 1.89  & $ 1.82 \pm 0.62 \times 10^{40} $ & $ 3.26 \pm 1.11 \times 10^{40} $ \\ 
27 & 4.39 & 1.8  & $ 80^{+ 31 }_{ -24 } $  & $ 2.55 \pm 0.67 \times 10^{43} $ & $ 2.32 \pm 0.61 \times 10^{44} $ \\ 
29 & 2.26 & 1.8  & $ 32^{+ 41 }_{ -19 } $  & $ 7.75 \pm 2.88 \times 10^{42} $ & $ 2.24 \pm 0.83 \times 10^{43} $ \\ 
31 & 0.86 & 1.8  & $ 6.7^{+ 5.9 }_{ -3.2 } $  & $ 2.41 \pm 0.94 \times 10^{42} $ & $ 4.40 \pm 1.71 \times 10^{42} $ \\ 
34 & 2.94 & 1.7 $ \pm $ 0.4 &  $ 30 ^{+ 32 }_{ -16 } $ & $ 4.14 \pm 0.59 \times 10^{43} $ & $ 1.74 \pm 0.25 \times 10^{44} $ \\ 
39 & 2.81 & 1.0 $ \pm $ 0.5 & $<$ 8.8  & $ 3.03 \pm 0.44 \times 10^{42} $ & $ 2.83 \pm 0.41 \times 10^{43} $ \\ 
45 & 0.74 & 1.8  & $<$ 0.1    & $ 1.19 \pm 0.16 \times 10^{41} $ & $ 2.22 \pm 0.31 \times 10^{41} $ \\ 
48 & 3.20 & 1.1 $ \pm $ 0.3 &  $ 23 ^{+ 36 }_{ -14 } $ & $ 4.93 \pm 0.75 \times 10^{42} $ & $ 4.64 \pm 0.70 \times 10^{43} $ \\ 
57 & 0.56 & 1.5 $ \pm $ 0.3 &  $ 3.8 ^{+ 2.4 }_{ -1.8 } $ & $ 7.88 \pm 1.10 \times 10^{41} $ & $ 2.03 \pm 0.28 \times 10^{42} $ \\ 
63 & 0.67 & 1.4 $ \pm $ 0.1 & $<$ 0.38  & $ 1.20 \pm 0.11 \times 10^{42} $ & $ 3.21 \pm 0.29 \times 10^{42} $ \\ 
67 & 1.99 & 1.6 $ \pm $ 0.1 &  $ 5.0 ^{+ 0.7 }_{ -0.6 } $ & $ 5.14 \pm 0.13 \times 10^{43} $ & $ 1.62 \pm 0.04 \times 10^{44} $ \\ 
81 & 0.98 & 1.8  & $<$ 2.13  & $ 2.18 \pm 1.66 \times 10^{41} $ & $ 3.83 \pm 2.92 \times 10^{41} $ \\ 
83 & 1.50 & 1.0 $ \pm $ 0.2 &  $ 14 ^{+ 1 }_{ -1 } $ & $ 1.76 \pm 0.04 \times 10^{43} $ & $ 1.13 \pm 0.03 \times 10^{44} $ \\ 
97 & 0.57 & 1.7 $ \pm $ 0.1 &  $ 6.7 ^{+ 0.6 }_{ -0.6 } $ & $ 6.44 \pm 0.20 \times 10^{42} $ & $ 1.22 \pm 0.04 \times 10^{43} $ \\ 
109 & 0.72 & 1.0 $ \pm $ 0.7 &  $ 15 ^{+ 6 }_{ -3 } $ & $ 2.03 \pm 0.24 \times 10^{42} $ & $ 1.09 \pm 0.13 \times 10^{43} $ \\ 
114 & 1.00 & 1.8  & $<$ 0.0  & $ 7.30 \pm 1.09 \times 10^{40} $ & $ 1.29 \pm 0.19 \times 10^{41} $ \\ 
115 & 1.07 & 1.8  & $ 9.1^{+ 11.7 }_{ -7.0 } $  & $ 7.32 \pm 1.03 \times 10^{41} $ & $ 1.33 \pm 0.19 \times 10^{42} $ \\ 
120 & 1.02 & 1.8  & $<$ 0.01  & $ 1.37 \pm 0.52 \times 10^{41} $ & $ 3.44 \pm 1.30 \times 10^{41} $ \\ 
127 & 0.77 & 1.8  & $<$ 0.01  & $ 1.06 \pm 0.30 \times 10^{42} $ & $ 1.73 \pm 0.48 \times 10^{42} $ \\ 
131 & 0.03 & 1.8  & $ 14^{+ 27 }_{ -10 } $  & $ 1.30 \pm 0.29 \times 10^{39} $ & $ 2.36 \pm 0.52 \times 10^{39} $ \\ 
132 & 3.53 & 1.8  & $ 13^{+ 9 }_{ -8 } $  & $ 4.63 \pm 1.27 \times 10^{42} $ & $ 2.36 \pm 0.65 \times 10^{43} $ \\ 
139 & 1.00 & 1.0 $ \pm $ 0.5 &  $ 1.8 ^{+ 1.2 }_{ -0.9 } $ & $ 1.38 \pm 0.19 \times 10^{42} $ & $ 7.91 \pm 1.06 \times 10^{42} $ \\ 
140 & 0.31 & 1.8  & $<$ 0.09  & $ 3.78 \pm 0.52 \times 10^{40} $ & $ 1.06 \pm 0.15 \times 10^{41} $ \\ 
141 & 0.54 & 1.9  & $<$ 0.01  & $ 4.79 \pm 0.03 \times 10^{43} $ & $ 6.86 \pm 0.04 \times 10^{43} $ \\ 
151 & 0.73 & 2.5 $ \pm $ 0.2 & $<$ 0.07  & $ 1.71 \pm 0.19 \times 10^{42} $ & $ 1.04 \pm 0.12 \times 10^{42} $ \\ 
152 & 0.74 & 1.8  & $<$ 0.81  & $ 1.27 \pm 0.17 \times 10^{41} $ & $ 2.06 \pm 0.28 \times 10^{41} $ \\ 
154 & 0.08 & 1.8  & $<$ 0.11  & $ 1.35 \pm 0.58 \times 10^{39} $ & $ 2.12 \pm 0.91 \times 10^{39} $ \\ 
156 & 1.16 & 1.8  & $ 2.9^{+ 1.5 }_{ -1.2 } $  & $ 1.46 \pm 0.20 \times 10^{42} $ & $ 2.89 \pm 0.39 \times 10^{42} $ \\ 
160 & 1.26 & 1.8 $ \pm $ 0.0 &  $ 0.94 ^{+ 0.15 }_{ -0.15 } $ & $ 4.66 \pm 0.08 \times 10^{43} $ & $ 8.70 \pm 0.14 \times 10^{43} $ \\ 
161 & 0.65 & 1.8  & $<$ 5.26  & $ 2.32 \pm 1.59 \times 10^{40} $ & $ 9.45 \pm 6.45 \times 10^{40} $ \\ 
163 & 1.61 & 1.8  & $<$ 0.78  & $ 9.62 \pm 2.84 \times 10^{41} $ & $ 2.25 \pm 0.66 \times 10^{42} $ \\ 
166 & 1.62 & 1.7 $ \pm $ 0.0 & $<$ 0.12  & $ 5.11 \pm 0.08 \times 10^{43} $ & $ 1.31 \pm 0.02 \times 10^{44} $ \\ 
164 & 0.42 & 1.8  & $<$ 0.35  & $ 2.89 \pm 1.15 \times 10^{40} $ & $ 5.92 \pm 2.36 \times 10^{40} $ \\ 
168 & 0.18 & 1.4 $ \pm $ 0.1 & $<$ 0.07  & $ 9.34 \pm 0.55 \times 10^{40} $ & $ 2.81 \pm 0.16 \times 10^{41} $ \\ 
174 & 0.58 & 1.8  & $ 1.1^{+ 1.6 }_{ -0.9 } $  & $ 1.06 \pm 0.42 \times 10^{41} $ & $ 1.81 \pm 0.72 \times 10^{41} $ \\ 
180 & 1.03 & 1.4 $ \pm $ 0.1 &  $ 2.6 ^{+ 0.8 }_{ -0.8 } $ & $ 2.82 \pm 0.14 \times 10^{42} $ & $ 8.86 \pm 0.44 \times 10^{42} $ \\ 
193 & 0.61 & 1.6 $ \pm $ 0.1 &  $ 2.2 ^{+ 0.3 }_{ -0.3 } $ & $ 6.43 \pm 0.15 \times 10^{42} $ & $ 1.43 \pm 0.03 \times 10^{43} $ \\ 
197 & 2.68 & 1.6 $ \pm $ 0.3 &  $ 21 ^{+ 18 }_{ -14 } $ & $ 1.46 \pm 0.21 \times 10^{43} $ & $ 6.32 \pm 0.92 \times 10^{43} $ \\ 
198 & 0.74 & 1.0 $ \pm $ 0.5 &  $ 18 ^{+ 7 }_{ -3 } $ & $ 1.48 \pm 0.12 \times 10^{42} $ & $ 8.15 \pm 0.65 \times 10^{42} $ \\ 
202 & 2.08 & 1.8  & $ 5.5^{+ 4.8 }_{ -3.4 } $  & $ 1.55 \pm 0.37 \times 10^{42} $ & $ 4.11 \pm 0.99 \times 10^{42} $ \\ 
217 & 0.24 & 1.8  & $ 0.65^{+ 0.27 }_{ -0.23 } $  & $ 4.87 \pm 0.60 \times 10^{40} $ & $ 8.03 \pm 1.00 \times 10^{40} $ \\ 
226 & 1.41 & 1.8  & $ 33^{+ 46 }_{ -21 } $  & $ 1.25 \pm 0.47 \times 10^{42} $ & $ 3.21 \pm 1.22 \times 10^{42} $ \\ 
228 & 0.66 & 1.8  & $<$ 0.09  & $ 1.41 \pm 0.23 \times 10^{41} $ & $ 3.85 \pm 0.62 \times 10^{41} $ \\ 
229 & 1.32 & 2.0 $ \pm $ 0.0 & $<$ 0.09  & $ 1.02 \pm 0.02 \times 10^{44} $ & $ 1.55 \pm 0.02 \times 10^{44} $ \\ 
241 & 0.57 & 1.8 $ \pm $ 0.0 & $<$ 0.02  & $ 5.13 \pm 0.10 \times 10^{42} $ & $ 7.49 \pm 0.15 \times 10^{42} $ \\ 
243 & 1.10 & 1.6 $ \pm $ 0.1 &  $ 19 ^{+ 2 }_{ -2 } $ & $ 2.87 \pm 0.08 \times 10^{43} $ & $ 7.44 \pm 0.20 \times 10^{43} $ \\ 
244 & 2.28 & 1.8  & $<$ 2.11  & $ 5.32 \pm 0.79 \times 10^{41} $ & $ 1.53 \pm 0.23 \times 10^{42} $ \\ 
249 & 0.12 & 1.8  & $<$ 0.65  & $ 3.02 \pm 0.81 \times 10^{39} $ & $ 4.74 \pm 1.27 \times 10^{39} $ \\ 
250 & 0.73 & 1.8  & $<$ 0.03  & $ 3.42 \pm 0.46 \times 10^{41} $ & $ 5.51 \pm 0.75 \times 10^{41} $ \\ 
251 & 0.67 & 1.8  & $<$ 0.01  & $ 1.53 \pm 0.43 \times 10^{42} $ & $ 2.55 \pm 0.72 \times 10^{42} $ \\ 
257 & 1.54 & 1.0 $ \pm $ 0.3 &  $ 28 ^{+ 2 }_{ 8 } $ & $ 6.75 \pm 0.29 \times 10^{42} $ & $ 4.49 \pm 0.19 \times 10^{43} $ \\ 
262 & 3.66 & 1.0 $ \pm $ 0.3 &  $ 41 ^{+ 14 }_{ -9 } $ & $ 3.53 \pm 0.31 \times 10^{42} $ & $ 4.85 \pm 0.43 \times 10^{43} $ \\ 
271 & 0.76 & 1.8  & $<$ 0.01 & $ 1.92 \pm 0.65 \times 10^{41} $ & $ 3.41 \pm 1.12 \times 10^{41} $ \\ 
273 & 1.98 & 1.8  & $ 6.9^{+ 8.1 }_{ -5.9 } $  & $ 1.38 \pm 0.20 \times 10^{42} $ & $ 3.53 \pm 0.50 \times 10^{42} $ \\ 
274 & 0.23 & 1.8  & $<$ 0.10  & $ 3.23 \pm 0.62 \times 10^{40} $ & $ 7.74 \pm 1.48 \times 10^{40} $ \\ 
279 & 1.22 & 1.8  & $<$ 0.10  & $ 9.49 \pm 4.74 \times 10^{40} $ & $ 2.64 \pm 1.32 \times 10^{41} $ \\ 
287 & 0.68 & 1.8  & $<$ 0.95  & $ 1.02 \pm 0.47 \times 10^{41} $ & $ 1.64 \pm 0.76 \times 10^{41} $ \\ 
\hline
\end{tabular}
\end{minipage}
\end{table*}

\begin{table*}
 \centering
 \begin{minipage}{120mm}
  \contcaption{}
  \begin{tabular}{@{}rrrrrrr@{}}
  \hline
  \hline
Xue X ID & z & $\Gamma$ & $N_H$ ($10^{22}$ cm$^{-2}$) & $L_{0.5-2keV}$ erg s$^{-1}$ &  $L_{2-10keV}$ erg s$^{-1}$ \\
  \hline
289 & 0.67 & 1.8  & $<$ 0.32  & $ 6.50 \pm 2.89 \times 10^{40} $ & $ 1.19 \pm 0.53 \times 10^{41} $ \\ 
290 & 2.83 & 1.5 $ \pm $ 0.3 &  $ 63 ^{+ 43 }_{ -33 } $ & $ 1.21 \pm 0.13 \times 10^{43} $ & $ 6.42 \pm 0.69 \times 10^{43} $ \\ 
292 & 0.52 & 1.8  & $<$ 0.06  & $ 1.00 \pm 0.24 \times 10^{41} $ & $ 1.76 \pm 0.42 \times 10^{41} $ \\ 
299 & 2.12 & 1.8  & $<$ 3.56  & $ 9.07 \pm 1.22 \times 10^{41} $ & $ 2.91 \pm 0.39 \times 10^{42} $ \\ 
300 & 0.20 & 1.8  & $ 0.37^{+ 0.37 }_{ -0.26 } $  & $ 2.29 \pm 0.62 \times 10^{40} $ & $ 3.68 \pm 1.00 \times 10^{40} $ \\ 
305 & 2.62 & 1.8  & $ 2.9^{+ 3.0 }_{ -2.5 } $  & $ 3.42 \pm 0.51 \times 10^{42} $ & $ 1.16 \pm 0.17 \times 10^{43} $ \\ 
311 & 0.74 & 1.8  & $<$ 0.36  & $ 1.69 \pm 0.31 \times 10^{41} $ & $ 5.06 \pm 0.92 \times 10^{41} $ \\ 
314 & 1.64 & 1.8  & $<$ 1.68  & $ 1.33 \pm 0.43 \times 10^{42} $ & $ 3.74 \pm 1.22 \times 10^{42} $ \\ 
316 & 0.66 & 1.8  & $<$ 1.82  & $ 5.08 \pm 0.65 \times 10^{40} $ & $ 1.13 \pm 0.14 \times 10^{41} $ \\ 
321 & 1.03 & 1.8  & $ 305^{+ 162 }_{ -115 } $  & $ 1.75 \pm 0.78 \times 10^{43} $ & $ 3.15 \pm 1.41 \times 10^{43} $ \\ 
322 & 0.58 & 1.8  & $<$ 0.85  & $ 1.04 \pm 0.18 \times 10^{41} $ & $ 1.63 \pm 0.28 \times 10^{41} $ \\ 
324 & 0.12 & 1.8  & $<$ 1.95  & $ 1.75 \pm 0.55 \times 10^{40} $ & $ 3.41 \pm 1.07 \times 10^{40} $ \\ 
327 & 0.53 & 1.7 $ \pm $ 0.1 & $<$ 0.06  & $ 1.81 \pm 0.06 \times 10^{42} $ & $ 3.43 \pm 0.12 \times 10^{42} $ \\ 
329 & 1.05 & 1.0 $ \pm $ 1.0 &  $ 12 ^{+ 141 }_{ -5 } $ & $ 1.29 \pm 0.27 \times 10^{42} $ & $ 7.85 \pm 1.61 \times 10^{42} $ \\ 
330 & 0.58 & 1.8  & $ 0.96^{+ 0.89 }_{ -0.73 } $  & $ 1.46 \pm 0.55 \times 10^{41} $ & $ 2.72 \pm 1.02 \times 10^{41} $ \\ 
343 & 1.09 & 2.3 $ \pm $ 0.5 &  $ 48 ^{+ 28 }_{ -26 } $ & $ 2.49 \pm 0.35 \times 10^{43} $ & $ 2.33 \pm 0.32 \times 10^{43} $ \\ 
344 & 1.62 & 1.9 $ \pm $ 0.1 &  $ 0.33 ^{+ 0.34 }_{ -0.31 } $ & $ 2.85 \pm 0.08 \times 10^{43} $ & $ 5.59 \pm 0.15 \times 10^{43} $ \\ 
348 & 2.54 & 1.8  & $<$ 2.03  & $ 1.35 \pm 0.45 \times 10^{42} $ & $ 4.31 \pm 1.44 \times 10^{42} $ \\ 
349 & 0.90 & 1.8  & $<$ 0.32  & $ 1.71 \pm 0.29 \times 10^{41} $ & $ 3.95 \pm 0.66 \times 10^{41} $ \\ 
367 & 1.03 & 2.3 $ \pm $ 0.0 & $<$ 0.01  & $ 1.10 \pm 0.01 \times 10^{44} $ & $ 9.44 \pm 0.11 \times 10^{43} $ \\ 
383 & 1.10 & 1.8  & $<$ 0.60  & $ 1.61 \pm 0.56 \times 10^{41} $ & $ 6.12 \pm 2.11 \times 10^{41} $ \\ 
385 & 0.25 & 1.8  & $<$ 1.34  & $ 1.23 \pm 0.21 \times 10^{40} $ & $ 2.74 \pm 0.46 \times 10^{40} $ \\ 
387 & 0.21 & 1.8  & $<$ 0.10  & $ 4.49 \pm 1.56 \times 10^{40} $ & $ 7.04 \pm 2.44 \times 10^{40} $ \\ 
391 & 0.74 & 1.8 $ \pm $ 0.1 & $<$ 0.12  & $ 6.42 \pm 0.51 \times 10^{41} $ & $ 1.05 \pm 0.08 \times 10^{42} $ \\ 
393 & 0.22 & 1.7 $ \pm $ 0.1 &  $ 0.27 ^{+ 0.14 }_{ -0.07 } $ & $ 8.31 \pm 0.46 \times 10^{40} $ & $ 1.64 \pm 0.09 \times 10^{41} $ \\ 
396 & 0.62 & 1.5 $ \pm $ 0.1 & $<$ 0.11  & $ 2.34 \pm 0.11 \times 10^{42} $ & $ 5.99 \pm 0.28 \times 10^{42} $ \\ 
407 & 0.64 & 1.8  & $ 0.71^{+ 0.71 }_{ -0.56 } $  & $ 3.52 \pm 0.59 \times 10^{41} $ & $ 6.13 \pm 1.03 \times 10^{41} $ \\ 
413 & 0.08 & 2.7 $ \pm $ 0.1 & $<$ 0.01  & $ 1.10 \pm 0.06 \times 10^{40} $ & $ 4.67 \pm 0.26 \times 10^{39} $ \\ 
418 & 0.08 & 2.1 $ \pm $ 0.1 & $<$ 0.01  & $ 8.51 \pm 0.52 \times 10^{39} $ & $ 8.30 \pm 0.51 \times 10^{39} $ \\ 
420 & 0.96 & 2.0 $ \pm $ 0.1 & $<$ 0.06  & $ 5.65 \pm 0.19 \times 10^{42} $ & $ 8.09 \pm 0.27 \times 10^{42} $ \\ 
429 & 1.04 & 1.0 $ \pm $ 2.9 &  $<$ 0.10  & $ 3.87 \pm 0.66 \times 10^{41} $ & $ 3.05 \pm 0.51 \times 10^{42} $ \\ 
435 & 2.22 & 1.8  & $<$ 0.38  & $ 1.72 \pm 0.32 \times 10^{42} $ & $ 4.82 \pm 0.89 \times 10^{42} $ \\ 
436 & 1.61 & 1.8  & $ 302^{+ 268 }_{ -141 } $  & $ 1.36 \pm 0.63 \times 10^{43} $ & $ 3.03 \pm 1.40 \times 10^{43} $ \\ 
453 & 0.67 & 1.8  & $<$ 0.16  & $ 1.38 \pm 0.46 \times 10^{41} $ & $ 2.46 \pm 0.82 \times 10^{41} $ \\ 
460 & 3.62 & 1.8  & $ 14^{+ 67 }_{ -14 } $  & $ 2.02 \pm 0.64 \times 10^{42} $ & $ 1.18 \pm 0.37 \times 10^{43} $ \\ 
467 & 1.38 & 1.8  & $ 12^{+ 29 }_{ -11 } $  & $ 4.79 \pm 3.20 \times 10^{41} $ & $ 1.12 \pm 0.75 \times 10^{42} $ \\ 
469 & 3.00 & 1.7 $ \pm $ 0.4 &  $ 27 ^{+ 22 }_{ -13 } $ & $ 1.58 \pm 0.28 \times 10^{43} $ & $ 6.73 \pm 1.18 \times 10^{43} $ \\ 
473 & 0.66 & 2.0 $ \pm $ 0.1 &  $ 3.7 ^{+ 0.5 }_{ -0.4 } $ & $ 2.97 \pm 0.09 \times 10^{43} $ & $ 3.28 \pm 0.10 \times 10^{43} $ \\ 
474 & 1.97 & 1.8  & $<$ 9.1  & $ 6.91 \pm 1.27 \times 10^{41} $ & $ 2.61 \pm 0.48 \times 10^{42} $ \\ 
476 & 1.11 & 1.8  & $<$ 1.57  & $ 1.78 \pm 0.92 \times 10^{41} $ & $ 3.29 \pm 1.70 \times 10^{41} $ \\ 
483 & 0.04 & 1.8  & $<$ 0.15  & $ 4.93 \pm 0.79 \times 10^{38} $ & $ 7.73 \pm 1.24 \times 10^{38} $ \\ 
490 & 2.58 & 1.8  & $ 61^{+ 35 }_{ -18 } $  & $ 7.45 \pm 1.35 \times 10^{42} $ & $ 2.81 \pm 0.51 \times 10^{43} $ \\ 
500 & 0.55 & 1.8  & $<$ 0.14  & $ 1.38 \pm 0.21 \times 10^{41} $ & $ 2.43 \pm 0.38 \times 10^{41} $ \\ 
499 & 1.61 & 1.6 $ \pm $ 0.1 & $<$ 1.45  & $ 2.22 \pm 0.19 \times 10^{42} $ & $ 7.46 \pm 0.65 \times 10^{42} $ \\ 
510 & 0.52 & 1.8  & $<$ 1.01  & $ 7.03 \pm 2.31 \times 10^{40} $ & $ 1.69 \pm 0.55 \times 10^{41} $ \\ 
507 & 0.13 & 1.8  & $<$ 0.50  & $ 5.56 \pm 2.50 \times 10^{39} $ & $ 2.32 \pm 1.05 \times 10^{40} $ \\ 
517 & 1.62 & 1.8  & $<$ 1.90  & $ 4.20 \pm 1.64 \times 10^{41} $ & $ 1.20 \pm 0.47 \times 10^{42} $ \\ 
518 & 1.60 & 2.1 $ \pm $ 0.0 &  $ 2.0 ^{+ 0.2 }_{ -0.2 } $ & $ 1.23 \pm 0.02 \times 10^{44} $ & $ 1.75 \pm 0.03 \times 10^{44} $ \\ 
519 & 2.44 & 1.8  & $ 33^{+ 20 }_{ -15 } $  & $ 5.86 \pm 2.62 \times 10^{42} $ & $ 1.80 \pm 0.80 \times 10^{43} $ \\ 
521 & 3.42 & 1.8  & $<$ 29.3  & $ 2.07 \pm 0.28 \times 10^{42} $ & $ 9.97 \pm 1.37 \times 10^{42} $ \\ 
530 & 0.74 & 1.8  & $<$ 0.28  & $ 1.00 \pm 0.15 \times 10^{41} $ & $ 2.09 \pm 0.31 \times 10^{41} $ \\ 
531 & 0.24 & 1.8  & $<$ 1.56  & $ 7.49 \pm 2.42 \times 10^{39} $ & $ 1.18 \pm 0.38 \times 10^{40} $ \\ 
533 & 0.30 & 1.5 $ \pm $ 0.1 &  $ 3.1 ^{+ 0.2 }_{ -0.2 } $ & $ 4.93 \pm 0.08 \times 10^{42} $ & $ 1.23 \pm 0.02 \times 10^{43} $ \\ 
542 & 0.69 & 1.8  & $ 1.6^{+ 2.0 }_{ -1.3 } $  & $ 2.39 \pm 0.84 \times 10^{41} $ & $ 4.30 \pm 1.52 \times 10^{41} $ \\ 
546 & 3.06 & 1.5 $ \pm $ 0.1 &  $ 42 ^{+ 5 }_{ -5 } $ & $ 5.00 \pm 0.16 \times 10^{43} $ & $ 2.81 \pm 0.09 \times 10^{44} $ \\ 
556 & 0.62 & 1.0 $ \pm $ 0.2 &  $ 2.3 ^{+ 0.4 }_{ -0.2 } $ & $ 1.71 \pm 0.07 \times 10^{42} $ & $ 9.13 \pm 0.38 \times 10^{42} $ \\ 
566 & 0.15 & 1.8  & $<$ 0.15  & $ 2.04 \pm 0.73 \times 10^{40} $ & $ 5.53 \pm 1.97 \times 10^{40} $ \\ 
569 & 0.66 & 1.8  & $ 0.80^{+ 0.69 }_{ -0.53 } $  & $ 8.37 \pm 2.03 \times 10^{41} $ & $ 1.35 \pm 0.33 \times 10^{42} $ \\ 
582 & 0.28 & 2.8 $ \pm $ 0.8 & $<$ 0.28  & $ 4.48 \pm 0.44 \times 10^{40} $ & $ 1.60 \pm 0.16 \times 10^{40} $ \\ 
590 & 0.41 & 2.0 $ \pm $ 0.2 & $<$ 0.11  & $ 1.70 \pm 0.15 \times 10^{41} $ & $ 2.14 \pm 0.19 \times 10^{41} $ \\ 
595 & 0.71 & 1.8  & $<$ 0.40  & $ 1.78 \pm 0.75 \times 10^{41} $ & $ 3.82 \pm 1.61 \times 10^{41} $ \\ 
596 & 0.21 & 1.8  & $<$ 0.14  & $ 6.24 \pm 2.00 \times 10^{39} $ & $ 1.20 \pm 0.38 \times 10^{40} $ \\ 
\hline
\end{tabular}
\end{minipage}
\end{table*}

\begin{table*}
 \centering
 \begin{minipage}{120mm}
  \contcaption{}
  \begin{tabular}{@{}rrrrrrr@{}}
  \hline
  \hline
Xue X ID & z & $\Gamma$ & $N_H$ ($10^{22}$ cm$^{-2}$) & $L_{0.5-2keV}$ erg s$^{-1}$ &  $L_{2-10keV}$ erg s$^{-1}$ \\
  \hline
597 & 0.46 & 1.8  & $<$ 0.08  & $ 6.67 \pm 0.76 \times 10^{40} $ & $ 1.20 \pm 0.14 \times 10^{41} $ \\ 
598 & 0.44 & 1.8  & $<$ 0.18  & $ 4.44 \pm 0.61 \times 10^{40} $ & $ 6.96 \pm 0.95 \times 10^{40} $ \\ 
609 & 0.10 & 1.8  & $<$ 0.07  & $ 8.63 \pm 0.93 \times 10^{39} $ & $ 2.08 \pm 0.22 \times 10^{40} $ \\ 
612 & 1.99 & 1.8  & $ 44^{+ 19 }_{ -17 } $  & $ 6.54 \pm 3.72 \times 10^{42} $ & $ 1.83 \pm 1.04 \times 10^{43} $ \\ 
617 & 1.10 & 1.8  & $ 6.5^{+ 2.4 }_{ -2.1 } $  & $ 5.27 \pm 0.98 \times 10^{41} $ & $ 1.01 \pm 0.19 \times 10^{42} $ \\ 
623 & 0.73 & 1.8  & $<$ 0.40  & $ 3.83 \pm 0.96 \times 10^{41} $ & $ 6.15 \pm 1.54 \times 10^{41} $ \\ 
626 & 0.98 & 1.7 $ \pm $ 0.0 &  $ 1.8 ^{+ 0.1 }_{ -0.1 } $ & $ 6.73 \pm 0.08 \times 10^{43} $ & $ 1.28 \pm 0.01 \times 10^{44} $ \\ 
629 & 0.67 & 1.6 $ \pm $ 0.1 &  $ 2.1 ^{+ 0.6 }_{ -0.5 } $ & $ 1.89 \pm 0.11 \times 10^{42} $ & $ 4.16 \pm 0.24 \times 10^{42} $ \\ 
631 & 1.12 & 1.8  & $<$ 0.89  & $ 3.26 \pm 0.41 \times 10^{41} $ & $ 6.49 \pm 0.81 \times 10^{41} $ \\ 
634 & 0.55 & 1.0 $ \pm $ 0.8 &  $ 4.1 ^{+ 2.9 }_{ -1.3 } $ & $ 7.00 \pm 1.00 \times 10^{41} $ & $ 3.83 \pm 0.55 \times 10^{42} $ \\ 
638 & 0.34 & 1.8  & $<$ 0.09  & $ 3.51 \pm 0.84 \times 10^{40} $ & $ 5.77 \pm 1.38 \times 10^{40} $ \\ 
644 & 0.97 & 1.8  & $<$ 0.72  & $ 1.41 \pm 0.32 \times 10^{42} $ & $ 2.48 \pm 0.55 \times 10^{42} $ \\ 
652 & 1.02 & 2.3 $ \pm $ 0.3 &  $ 1.5 ^{+ 1.4 }_{ -0.9 } $ & $ 5.12 \pm 0.56 \times 10^{42} $ & $ 4.77 \pm 0.52 \times 10^{42} $ \\ 
654 & 0.53 & 1.8  & $<$ 0.30  & $ 9.49 \pm 3.60 \times 10^{40} $ & $ 1.49 \pm 0.56 \times 10^{41} $ \\ 
656 & 1.37 & 1.8  & $ 16^{+ 11 }_{ -7 } $  & $ 1.80 \pm 0.96 \times 10^{42} $ & $ 3.64 \pm 1.95 \times 10^{42} $ \\ 
658 & 0.98 & 1.7 $ \pm $ 0.4 &  $ 16 ^{+ 1 }_{ 8 } $ & $ 4.33 \pm 0.98 \times 10^{42} $ & $ 8.63 \pm 1.95 \times 10^{42} $ \\ 
671 & 0.37 & 1.8  & $<$ 0.28  & $ 8.50 \pm 2.65 \times 10^{40} $ & $ 1.47 \pm 0.46 \times 10^{41} $ \\ 
681 & 0.73 & 1.8 $ \pm $ 0.1 & $<$ 0.04  & $ 6.74 \pm 0.20 \times 10^{42} $ & $ 1.03 \pm 0.03 \times 10^{43} $ \\ 
682 & 1.15 & 1.8  & $ 1.8^{+ 4.2 }_{ -1.8 } $  & $ 6.49 \pm 0.76 \times 10^{41} $ & $ 1.33 \pm 0.16 \times 10^{42} $ \\ 
683 & 1.20 & 2.0 $ \pm $ 0.1 &  $ 17 ^{+ 2 }_{ -1 } $ & $ 4.07 \pm 0.16 \times 10^{43} $ & $ 6.13 \pm 0.25 \times 10^{43} $ \\ 
695 & 0.62 & 1.7 $ \pm $ 0.1 & $<$ 0.04  & $ 2.10 \pm 0.09 \times 10^{42} $ & $ 4.09 \pm 0.17 \times 10^{42} $ \\ 
698 & 3.10 & 1.8 $ \pm $ 0.1 &  $ 2.9 ^{+ 1.1 }_{ -0.9 } $ & $ 8.64 \pm 0.23 \times 10^{43} $ & $ 3.55 \pm 0.09 \times 10^{44} $ \\ 
700 & 3.69 & 1.8 $ \pm $ 0.2 &  $ 3.5 ^{+ 6.6 }_{ -3.5 } $ & $ 1.97 \pm 0.27 \times 10^{43} $ & $ 1.12 \pm 0.15 \times 10^{44} $ \\ 
706 & 0.89 & 2.8 $ \pm $ 2.2 &  $ 93 ^{+ 31 }_{ -27 } $ & $ 1.01 \pm 0.31 \times 10^{44} $ & $ 4.19 \pm 1.30 \times 10^{43} $ \\ 
712 & 0.52 & 1.0 $ \pm $ 0.7 &  $ 0.81 ^{+ 1.42 }_{ -0.75 } $ & $ 4.04 \pm 0.71 \times 10^{41} $ & $ 2.31 \pm 0.41 \times 10^{42} $ \\ 
719 & 0.22 & 1.8  & $<$ 0.31  & $ 4.33 \pm 1.54 \times 10^{40} $ & $ 6.80 \pm 2.41 \times 10^{40} $ \\ 
722 & 0.73 & 2.8 $ \pm $ 1.0 &  $ 25 ^{+ 5 }_{ -6 } $ & $ 6.01 \pm 0.93 \times 10^{43} $ & $ 2.28 \pm 0.35 \times 10^{43} $ \\ 
726 & 0.52 & 1.5 $ \pm $ 0.2 &  $ 0.56 ^{+ 0.47 }_{ -0.38 } $ & $ 1.25 \pm 0.11 \times 10^{42} $ & $ 3.18 \pm 0.27 \times 10^{42} $ \\ 
727 & 0.18 & 1.8  & $<$ 0.14  & $ 2.63 \pm 1.15 \times 10^{40} $ & $ 4.24 \pm 1.86 \times 10^{40} $ \\ 
728 & 1.03 & 1.7 $ \pm $ 0.1 & $<$ 0.05  & $ 1.89 \pm 0.05 \times 10^{43} $ & $ 3.93 \pm 0.11 \times 10^{43} $ \\ 
731 & 0.56 & 1.8  & $<$ 0.75  & $ 2.63 \pm 0.60 \times 10^{41} $ & $ 5.06 \pm 1.16 \times 10^{41} $ \\ 
\hline
\end{tabular}
\end{minipage}
\end{table*}

\begin{table*}
 \centering
 \begin{minipage}{130mm}
  \caption{X--ray spectral analysis of sources with radio
  counterparts in the E-CDFS complementary area covered by the E-CDFS.
  X ID is the source ID from the catalog of \citet{lehmer05}.
  z is the spectroscopic or photometric redshift;  
  $\Gamma$ is the slope of the intrinsic
  X-ray emission modelled with a power law; $N_H$ is the column density; $L_{0.5-2keV}$
  and $L_{2-10keV}$ are the luminosities in the soft and X-ray band 
  respectively. \label{tab:table3}}
  \begin{tabular}{rrrrrrr}
  \hline
  \hline
Lehmer X ID & z & $\Gamma$ & $N_H$ ($10^{22}$ cm$^{-2}$) & $L_{0.5-2keV}$ erg s$^{-1}$ &  $L_{2-10keV}$ erg s$^{-1}$ \\
  \hline
7 & 1.37 & 2.0 $ \pm $ 0.1 & $ 0.71 ^{+ 0.30 }_{ -0.28 } $ & $ 2.65 \pm 0.09 \times 10^{44} $ & $ 4.27 \pm 0.14 \times 10^{44} $ \\ 
10 & 1.00 & 1.8  & $<$ 0.59  & $ 3.47 \pm 0.98 \times 10^{42} $ & $ 7.25 \pm 2.04 \times 10^{42} $ \\ 
26 & 1.00 & 2.4 $ \pm $ 0.3 & $ 0.71 ^{+ 0.67 }_{ -0.53 } $ & $ 5.67 \pm 0.60 \times 10^{43} $ & $ 4.32 \pm 0.45 \times 10^{43} $ \\ 
27 & 0.71 & 1.8  & $ 8.3 ^{+ 3.5 }_{ -4.2 } $ & $ 7.57 \pm 1.46 \times 10^{42} $ & $ 1.26 \pm 0.24 \times 10^{43} $ \\ 
29 & 0.54 & 1.8  & $ 2.6 ^{+ 1.2 }_{ -1.0 } $ & $ 2.61 \pm 0.61 \times 10^{42} $ & $ 4.28 \pm 1.00 \times 10^{42} $ \\ 
32 & 0.53 & 1.8  & $ 3.2 ^{+ 0.7 }_{ -0.6 } $ & $ 4.79 \pm 0.55 \times 10^{42} $ & $ 7.67 \pm 0.88 \times 10^{42} $ \\ 
45 & 3.41 & 1.8  & $ 5.2 ^{+ 4.4 }_{ -2.6 } $ & $ 8.38 \pm 0.92 \times 10^{43} $ & $ 4.19 \pm 0.46 \times 10^{44} $ \\ 
46 & 1.18 & 2.3 $ \pm $ 0.2 & $<$ 0.84  & $ 4.76 \pm 0.42 \times 10^{43} $ & $ 4.77 \pm 0.42 \times 10^{43} $ \\ 
47 & 0.54 & 2.1 $ \pm $ 0.2 & $ 5.5 ^{+ 0.9 }_{ -0.8 } $ & $ 3.86 \pm 0.21 \times 10^{43} $ & $ 4.08 \pm 0.22 \times 10^{43} $ \\ 
50 & 1.86 & 1.8  & $ 38 ^{+ 11 }_{ -10 } $ & $ 5.58 \pm 1.06 \times 10^{43} $ & $ 1.45 \pm 0.28 \times 10^{44} $ \\ 
51 & 4.50 & 1.8  & $<$ 6.69  & $ 1.71 \pm 0.35 \times 10^{43} $ & $ 1.72 \pm 0.36 \times 10^{44} $ \\ 
66 & 0.68 & 1.8  & $<$ 0.14  & $ 6.73 \pm 2.35 \times 10^{41} $ & $ 1.06 \pm 0.37 \times 10^{42} $ \\ 
72 & 0.69 & 1.8  & $ 0.53 ^{+ 0.35 }_{ -0.30 } $ & $ 8.97 \pm 0.97 \times 10^{42} $ & $ 1.46 \pm 0.16 \times 10^{43} $ \\ 
74 & 0.18 & 1.8  & $<$ 0.09  & $ 1.46 \pm 0.11 \times 10^{40} $ & $ 2.96 \pm 0.22 \times 10^{40} $ \\ 
75 & 0.75 & 1.8  & $ 2.7 ^{+ 1.8 }_{ -1.6 } $ & $ 4.20 \pm 1.14 \times 10^{42} $ & $ 8.08 \pm 2.19 \times 10^{42} $ \\ 
79 & 0.97 & 1.8  & $ 13 ^{+ 5 }_{ -3 } $ & $ 1.10 \pm 0.24 \times 10^{43} $ & $ 1.92 \pm 0.42 \times 10^{43} $ \\ 
82 & 0.14 & 1.8  & $<$ 0.07  & $ 1.12 \pm 0.39 \times 10^{40} $ & $ 2.13 \pm 0.75 \times 10^{40} $ \\ 
91 & 0.93 & 1.8  & $<$ 0.74  & $ 9.29 \pm 3.29 \times 10^{41} $ & $ 1.61 \pm 0.57 \times 10^{42} $ \\ 
93 & 0.66 & 1.8  & $ 15 ^{+ 10 }_{ -9 } $ & $ 2.55 \pm 0.81 \times 10^{42} $ & $ 4.00 \pm 1.27 \times 10^{42} $ \\ 
94 & 1.35 & 2.5 $ \pm $ 0.2 & $ 0.91 ^{+ 0.64 }_{ -0.55 } $ & $ 1.26 \pm 0.08 \times 10^{44} $ & $ 9.95 \pm 0.66 \times 10^{43} $ \\ 
104 & 2.47 & 1.8  & $ 43 ^{+ 18 }_{ -14 } $ & $ 2.77 \pm 0.74 \times 10^{43} $ & $ 8.57 \pm 2.28 \times 10^{43} $ \\ 
107 & 0.19 & 1.8  & $ 6.5 ^{+ 7.7 }_{ -2.5 } $ & $ 2.53 \pm 0.68 \times 10^{41} $ & $ 4.03 \pm 1.09 \times 10^{41} $ \\ 
118 & 0.47 & 1.5 $ \pm $ 0.1 & $ 0.87 ^{+ 0.23 }_{ -0.21 } $ & $ 8.37 \pm 0.39 \times 10^{42} $ & $ 2.05 \pm 0.10 \times 10^{43} $ \\ 
121 & 2.72 & 1.8  & $ 8.7 ^{+ 3.7 }_{ -3.9 } $ & $ 3.52 \pm 0.51 \times 10^{43} $ & $ 1.21 \pm 0.17 \times 10^{44} $ \\ 
124 & 1.00 & 1.8  & $ 4.2 ^{+ 0.8 }_{ -0.7 } $ & $ 2.04 \pm 0.21 \times 10^{43} $ & $ 3.62 \pm 0.37 \times 10^{43} $ \\ 
135 & 0.39 & 1.8  & $<$ 0.01  & $ 1.56 \pm 1.21 \times 10^{41} $ & $ 2.44 \pm 1.89 \times 10^{41} $ \\ 
136 & 1.27 & 1.5 $ \pm $ 0.3 & $ 7.1 ^{+ 2.6 }_{ -2.3 } $ & $ 2.40 \pm 0.21 \times 10^{43} $ & $ 7.76 \pm 0.68 \times 10^{43} $ \\ 
140 & 0.47 & 1.8  & $ 0.8 ^{+ 5.0 }_{ -0.7 } $ & $ 2.67 \pm 0.67 \times 10^{41} $ & $ 7.61 \pm 1.95 \times 10^{41} $ \\ 
141 & 0.31 & 1.8  & $ 1.1 ^{+ 1.1 }_{ -0.8 } $ & $ 2.71 \pm 1.37 \times 10^{41} $ & $ 4.25 \pm 2.15 \times 10^{41} $ \\ 
146 & 0.06 & 1.8  & $ 4.8 ^{+ 3.3 }_{ -2.2 } $ & $ 9.86 \pm 3.52 \times 10^{39} $ & $ 1.55 \pm 0.55 \times 10^{40} $ \\ 
148 & 1.00 & 1.8  & $ 0.87 ^{+ 0.96 }_{ -0.76 } $ & $ 5.04 \pm 2.29 \times 10^{42} $ & $ 9.58 \pm 4.36 \times 10^{42} $ \\ 
174 & 1.22 & 1.8  & $ 2.5 ^{+ 0.5 }_{ -0.8 } $ & $ 1.98 \pm 0.20 \times 10^{43} $ & $ 3.84 \pm 0.39 \times 10^{43} $ \\ 
177 & 1.00 & 1.8  & $ 2.8 ^{+ 5.5 }_{ -2.8 } $ & $ 1.59 \pm 0.71 \times 10^{42} $ & $ 2.83 \pm 1.26 \times 10^{42} $ \\ 
184 & 1.32 & 1.9 $ \pm $ 0.2 & $<$ 0.67  & $ 3.37 \pm 0.25 \times 10^{43} $ & $ 5.81 \pm 0.43 \times 10^{43} $ \\ 
188 & 1.28 & 1.8  & $ 23 ^{+ 6 }_{ -5 } $ & $ 3.02 \pm 0.43 \times 10^{43} $ & $ 5.95 \pm 0.84 \times 10^{43} $ \\ 
201 & 0.15 & 1.8  & $ 4.1 ^{+ 1.3 }_{ -1.0 } $ & $ 1.47 \pm 0.27 \times 10^{41} $ & $ 2.30 \pm 0.42 \times 10^{41} $ \\ 
205 & 2.06 & 1.8  & $<$ 9.5  & $ 2.88 \pm 1.05 \times 10^{42} $ & $ 1.22 \pm 0.44 \times 10^{43} $ \\ 
213 & 0.41 & 1.8  & $<$ 0.88  & $ 2.81 \pm 0.81 \times 10^{41} $ & $ 4.40 \pm 1.27 \times 10^{41} $ \\ 
222 & 0.68 & 1.8  & $ 4.3 ^{+ 3.1 }_{ -2.2 } $ & $ 2.05 \pm 0.45 \times 10^{42} $ & $ 3.24 \pm 0.71 \times 10^{42} $ \\ 
224 & 2.43 & 1.8  & $<$ 20.5  & $ 1.32 \pm 0.28 \times 10^{43} $ & $ 4.97 \pm 1.06 \times 10^{43} $ \\ 
235 & 1.38 & 1.8  & $<$ 0.10  & $ 8.13 \pm 1.47 \times 10^{42} $ & $ 1.66 \pm 0.30 \times 10^{43} $ \\ 
254 & 0.27 & 1.8  & $<$ 0.05  & $ 6.65 \pm 0.65 \times 10^{41} $ & $ 1.08 \pm 0.11 \times 10^{42} $ \\ 
265 & 1.01 & 1.2 $ \pm $ 0.4 & $ 13 ^{+ 5 }_{ -4 } $ & $ 1.59 \pm 0.16 \times 10^{43} $ & $ 6.37 \pm 0.63 \times 10^{43} $ \\ 
268 & 1.00 & 1.8  & $ 3.3 ^{+ 1.6 }_{ -1.4 } $ & $ 8.08 \pm 1.97 \times 10^{42} $ & $ 1.46 \pm 0.36 \times 10^{43} $ \\ 
278 & 0.96 & 1.8  & $ 6.3 ^{+ 2.8 }_{ -2.1 } $ & $ 1.19 \pm 0.44 \times 10^{43} $ & $ 2.16 \pm 0.79 \times 10^{43} $ \\ 
282 & 1.00 & 1.8  & $ 3.2 ^{+ 0.8 }_{ -0.7 } $ & $ 2.29 \pm 0.26 \times 10^{43} $ & $ 4.06 \pm 0.47 \times 10^{43} $ \\ 
289 & 0.96 & 1.7 $ \pm $ 0.2 & $ 0.70 ^{+ 0.61 }_{ -0.52 } $ & $ 3.04 \pm 0.22 \times 10^{43} $ & $ 6.43 \pm 0.47 \times 10^{43} $ \\ 
293 & 0.34 & 1.8  & $ 1.8 ^{+ 1.2 }_{ -1.1 } $ & $ 4.50 \pm 1.44 \times 10^{41} $ & $ 7.06 \pm 2.26 \times 10^{41} $ \\ 
307 & 1.94 & 1.8  & $ 28 ^{+ 7 }_{ -7 } $ & $ 4.97 \pm 0.73 \times 10^{43} $ & $ 1.26 \pm 0.19 \times 10^{44} $ \\ 
315 & 1.00 & 1.8  & $<$ 1.21  & $ 3.40 \pm 1.01 \times 10^{42} $ & $ 6.40 \pm 1.91 \times 10^{42} $ \\ 
319 & 3.53 & 1.8  & $ 19 ^{+ 8 }_{ -6 } $ & $ 8.00 \pm 1.32 \times 10^{43} $ & $ 4.09 \pm 0.68 \times 10^{44} $ \\ 
321 & 0.61 & 1.6 $ \pm $ 0.1 & $<$ 0.02  & $ 3.77 \pm 0.11 \times 10^{43} $ & $ 7.54 \pm 0.21 \times 10^{43} $ \\ 
330 & 3.16 & 1.8  & $ 24 ^{+ 14 }_{ -10 } $ & $ 1.38 \pm 0.28 \times 10^{44} $ & $ 5.82 \pm 1.18 \times 10^{44} $ \\ 
333 & 1.04 & 1.8  & $<$ 0.80  & $ 7.12 \pm 1.56 \times 10^{42} $ & $ 1.29 \pm 0.28 \times 10^{43} $ \\ 
379 & 0.73 & 1.9 $ \pm $ 0.0 & $ 0.11 ^{+ 0.03 }_{ -0.06 } $ & $ 1.26 \pm 0.02 \times 10^{44} $ & $ 1.76 \pm 0.03 \times 10^{44} $ \\ 
381 & 0.53 & 2.1 $ \pm $ 0.1 & $<$ 0.05  & $ 1.34 \pm 0.06 \times 10^{43} $ & $ 1.39 \pm 0.06 \times 10^{43} $ \\ 
385 & 0.53 & 1.9 $ \pm $ 0.1 & $ 0.27 ^{+ 0.07 }_{ -0.07 } $ & $ 8.27 \pm 0.18 \times 10^{43} $ & $ 1.08 \pm 0.02 \times 10^{44} $ \\ 
\hline
\end{tabular}
\end{minipage}
\end{table*}

\begin{table*}
 \centering
 \begin{minipage}{130mm}
  \contcaption{}
  \begin{tabular}{rrrrrrr}
  \hline
  \hline
Lehmer X ID & z & $\Gamma$ & $N_H$ ($10^{22}$ cm$^{-2}$) & $L_{0.5-2keV}$ erg s$^{-1}$ &  $L_{2-10keV}$ erg s$^{-1}$ \\
  \hline
398 & 1.95 & 1.8 $ \pm $ 0.1 & $<$ 0.13  & $ 3.27 \pm 0.11 \times 10^{44} $ & $ 8.32 \pm 0.28 \times 10^{44} $ \\ 
421 & 2.86 & 1.8  & $<$ 20.4  & $ 2.87 \pm 1.15 \times 10^{43} $ & $ 1.06 \pm 0.42 \times 10^{44} $ \\ 
419 & 0.65 & 1.1 $ \pm $ 0.2 & $<$ 0.20  & $ 3.00 \pm 0.25 \times 10^{42} $ & $ 1.38 \pm 0.12 \times 10^{43} $ \\ 
437 & 1.00 & 1.0 $ \pm $ 0.4 & $ 3.3 ^{+ 1.1 }_{ -0.7 } $ & $ 1.45 \pm 0.09 \times 10^{43} $ & $ 8.34 \pm 0.53 \times 10^{43} $ \\ 
445 & 0.28 & 1.8  & $<$ 0.44  & $ 8.15 \pm 3.11 \times 10^{40} $ & $ 1.28 \pm 0.49 \times 10^{41} $ \\ 
460 & 0.10 & 1.8  & $<$ 0.06  & $ 1.46 \pm 0.42 \times 10^{40} $ & $ 2.28 \pm 0.65 \times 10^{40} $ \\ 
461 & 0.15 & 1.8  & $<$ 0.11  & $ 1.29 \pm 0.49 \times 10^{40} $ & $ 2.43 \pm 0.92 \times 10^{40} $ \\ 
470 & 0.68 & 1.8  & $<$ 0.77  & $ 1.70 \pm 0.40 \times 10^{42} $ & $ 2.87 \pm 0.68 \times 10^{42} $ \\ 
504 & 0.15 & 1.8  & $<$ 0.02  & $ 2.88 \pm 0.62 \times 10^{40} $ & $ 5.08 \pm 1.10 \times 10^{40} $ \\ 
506 & 2.12 & 1.8  & $ 7.2 ^{+ 5.1 }_{ -3.4 } $ & $ 3.62 \pm 0.71 \times 10^{43} $ & $ 1.02 \pm 0.20 \times 10^{44} $ \\ 
508 & 0.27 & 1.8  & $<$ 0.19  & $ 1.05 \pm 0.21 \times 10^{41} $ & $ 1.65 \pm 0.33 \times 10^{41} $ \\ 
517 & 1.35 & 2.2 $ \pm $ 0.2 & $<$ 0.56  & $ 4.10 \pm 0.35 \times 10^{43} $ & $ 4.67 \pm 0.40 \times 10^{43} $ \\ 
532 & 2.01 & 1.9 $ \pm $ 0.1 & $<$ 0.64  & $ 1.79 \pm 0.08 \times 10^{44} $ & $ 4.09 \pm 0.18 \times 10^{44} $ \\ 
535 & 1.02 & 1.8  & $ 20 ^{+ 9 }_{ -6 } $ & $ 7.19 \pm 1.86 \times 10^{42} $ & $ 1.28 \pm 0.33 \times 10^{43} $ \\ 
538 & 0.20 & 2.2 $ \pm $ 0.2 & $ 2.3 ^{+ 0.5 }_{ -0.5 } $ & $ 2.20 \pm 0.15 \times 10^{42} $ & $ 2.02 \pm 0.14 \times 10^{42} $ \\ 
546 & 0.71 & 1.8  & $ 1.9 ^{+ 1.7 }_{ -1.5 } $ & $ 1.45 \pm 0.42 \times 10^{42} $ & $ 2.35 \pm 0.69 \times 10^{42} $ \\ 
552 & 0.62 & 1.8  & $<$ 0.27  & $ 2.55 \pm 0.33 \times 10^{42} $ & $ 4.05 \pm 0.53 \times 10^{42} $ \\ 
555 & 1.08 & 1.8  & $<$ 7.6  & $ 1.26 \pm 0.61 \times 10^{42} $ & $ 2.31 \pm 1.12 \times 10^{42} $ \\ 
557 & 0.30 & 1.8  & $<$ 0.10  & $ 4.39 \pm 0.30 \times 10^{40} $ & $ 6.88 \pm 0.47 \times 10^{40} $ \\ 
583 & 1.00 & 1.8  & $<$ 10  & $ 1.10 \pm 0.94 \times 10^{42} $ & $ 1.95 \pm 1.67 \times 10^{42} $ \\ 
584 & 2.07 & 1.8  & $ 3.0 ^{+ 1.2 }_{ -1.1 } $ & $ 4.95 \pm 0.45 \times 10^{43} $ & $ 1.35 \pm 0.12 \times 10^{44} $ \\ 
588 & 2.58 & 1.4 $ \pm $ 0.2 & $<$ 0.79  & $ 4.99 \pm 0.34 \times 10^{43} $ & $ 2.60 \pm 0.18 \times 10^{44} $ \\ 
587 & 0.67 & 1.8  & $<$ 3.52  & $ 2.96 \pm 1.85 \times 10^{41} $ & $ 4.64 \pm 2.90 \times 10^{41} $ \\ 
599 & 0.25 & 1.8  & $<$ 0.26  & $ 4.57 \pm 3.89 \times 10^{40} $ & $ 1.66 \pm 1.41 \times 10^{41} $ \\ 
600 & 1.30 & 1.8  & $ 6.1 ^{+ 2.3 }_{ -2.0 } $ & $ 1.74 \pm 0.43 \times 10^{43} $ & $ 3.67 \pm 0.91 \times 10^{43} $ \\ 
606 & 1.12 & 1.8  & $ 157 ^{+ 54 }_{ -51 } $ & $ 6.49 \pm 1.79 \times 10^{43} $ & $ 1.20 \pm 0.33 \times 10^{44} $ \\ 
607 & 1.18 & 1.8  & $<$ 3.25  & $ 2.63 \pm 0.66 \times 10^{42} $ & $ 4.99 \pm 1.24 \times 10^{42} $ \\ 
609 & 0.93 & 1.8  & $ 295 ^{+ 157 }_{ -92 } $ & $ 7.78 \pm 4.72 \times 10^{43} $ & $ 1.35 \pm 0.82 \times 10^{44} $ \\ 
611 & 2.56 & 2.0 $ \pm $ 0.1 & $<$ 0.32  & $ 2.13 \pm 0.10 \times 10^{44} $ & $ 5.65 \pm 0.28 \times 10^{44} $ \\ 
612 & 0.21 & 1.8  & $ 18 ^{+ 26 }_{ -8 } $ & $ 8.36 \pm 1.53 \times 10^{41} $ & $ 1.40 \pm 0.26 \times 10^{42} $ \\ 
616 & 1.21 & 1.8  & $ 5.1 ^{+ 1.9 }_{ -1.8 } $ & $ 1.53 \pm 0.20 \times 10^{43} $ & $ 3.09 \pm 0.41 \times 10^{43} $ \\ 
617 & 0.68 & 1.8  & $<$ 1.65  & $ 7.06 \pm 2.94 \times 10^{41} $ & $ 1.37 \pm 0.57 \times 10^{42} $ \\ 
623 & 0.50 & 1.8  & $<$ 1.63  & $ 3.14 \pm 2.50 \times 10^{41} $ & $ 5.21 \pm 4.14 \times 10^{41} $ \\ 
632 & 0.36 & 1.8  & $<$ 1.75  & $ 2.36 \pm 0.89 \times 10^{41} $ & $ 3.85 \pm 1.46 \times 10^{41} $ \\ 
634 & 0.09 & 1.8  & $<$ 0.26  & $ 1.56 \pm 0.52 \times 10^{40} $ & $ 2.45 \pm 0.81 \times 10^{40} $ \\ 
637 & 1.79 & 1.8  & $ 30 ^{+ 14 }_{ -7 } $ & $ 3.10 \pm 0.91 \times 10^{43} $ & $ 8.31 \pm 2.44 \times 10^{43} $ \\ 
639 & 0.15 & 1.8  & $<$ 0.21  & $ 9.20 \pm 4.30 \times 10^{39} $ & $ 1.47 \pm 0.69 \times 10^{40} $ \\ 
646 & 0.64 & 1.8  & $<$ 0.94  & $ 3.17 \pm 1.41 \times 10^{41} $ & $ 5.93 \pm 2.65 \times 10^{41} $ \\ 
655 & 1.03 & 1.8  & $ 26 ^{+ 12 }_{ -7 } $ & $ 1.99 \pm 0.39 \times 10^{43} $ & $ 3.69 \pm 0.72 \times 10^{43} $ \\ 
657 & 0.14 & 1.8  & $<$ 0.09  & $ 2.46 \pm 0.78 \times 10^{40} $ & $ 5.34 \pm 1.69 \times 10^{40} $ \\ 
658 & 0.81 & 1.8  & $ 3.5 ^{+ 1.1 }_{ -1.0 } $ & $ 7.18 \pm 1.44 \times 10^{42} $ & $ 1.25 \pm 0.25 \times 10^{43} $ \\ 
664 & 0.13 & 1.8  & $<$ 0.01  & $ 1.16 \pm 0.17 \times 10^{41} $ & $ 1.83 \pm 0.27 \times 10^{41} $ \\ 
669 & 0.13 & 2.8 $ \pm $ 0.7 & $<$ 0.02  & $ 2.55 \pm 0.28 \times 10^{41} $ & $ 9.10 \pm 0.99 \times 10^{40} $ \\ 
674 & 1.16 & 1.8  & $<$ 0.76  & $ 2.67 \pm 0.62 \times 10^{42} $ & $ 5.03 \pm 1.18 \times 10^{42} $ \\ 
680 & 1.39 & 1.8  & $ 1.9 ^{+ 1.4 }_{ -1.0 } $ & $ 2.00 \pm 0.24 \times 10^{43} $ & $ 4.10 \pm 0.50 \times 10^{43} $ \\ 
695 & 0.52 & 1.9 $ \pm $ 0.5 & $ 17 ^{+ 5 }_{ -4 } $ & $ 1.50 \pm 0.16 \times 10^{43} $ & $ 2.24 \pm 0.24 \times 10^{43} $ \\ 
697 & 0.45 & 1.8  & $ 1.1 ^{+ 0.4 }_{ -0.4 } $ & $ 2.28 \pm 0.40 \times 10^{42} $ & $ 3.72 \pm 0.66 \times 10^{42} $ \\ 
700 & 2.17 & 2.3 $ \pm $ 0.2 & $ 0.35 ^{+ 1.05 }_{ -0.35 } $ & $ 1.39 \pm 0.10 \times 10^{44} $ & $ 2.20 \pm 0.16 \times 10^{44} $ \\ 
711 & 2.49 & 1.8  & $ 11 ^{+ 7 }_{ -5 } $ & $ 3.30 \pm 0.69 \times 10^{43} $ & $ 1.08 \pm 0.22 \times 10^{44} $ \\ 
712 & 0.84 & 1.9 $ \pm $ 0.1 & $<$ 0.07  & $ 1.27 \pm 0.03 \times 10^{44} $ & $ 1.82 \pm 0.04 \times 10^{44} $ \\ 
716 & 0.76 & 2.2 $ \pm $ 0.1 & $<$ 0.09  & $ 4.85 \pm 0.20 \times 10^{43} $ & $ 4.36 \pm 0.18 \times 10^{43} $ \\ 
723 & 2.60 & 2.0 $ \pm $ 0.2 & $<$ 1.84  & $ 1.01 \pm 0.08 \times 10^{44} $ & $ 2.59 \pm 0.20 \times 10^{44} $ \\ 
725 & 1.31 & 2.0 $ \pm $ 0.1 & $<$ 0.16  & $ 1.35 \pm 0.05 \times 10^{44} $ & $ 2.08 \pm 0.08 \times 10^{44} $ \\ 
738 & 0.86 & 2.5 $ \pm $ 0.5 & $ 14 ^{+ 5 }_{ -4 } $ & $ 1.18 \pm 0.17 \times 10^{44} $ & $ 7.54 \pm 1.09 \times 10^{43} $ \\ 
743 & 0.51 & 1.8  & $ 2.4 ^{+ 1.2 }_{ -0.9 } $ & $ 1.73 \pm 0.27 \times 10^{42} $ & $ 2.72 \pm 0.42 \times 10^{42} $ \\ 
746 & 0.97 & 1.8  & $ 77 ^{+ 49 }_{ -38 } $ & $ 2.53 \pm 1.20 \times 10^{43} $ & $ 4.47 \pm 2.12 \times 10^{43} $ \\ 
748 & 1.15 & 1.8  & $<$ 1.72  & $ 5.06 \pm 1.76 \times 10^{42} $ & $ 9.68 \pm 3.37 \times 10^{42} $ \\ 
750 & 1.21 & 1.8  & $<$ 1.78  & $ 1.27 \pm 0.23 \times 10^{43} $ & $ 2.42 \pm 0.44 \times 10^{43} $ \\ 
752 & 3.34 & 1.8  & $<$ 5.45  & $ 4.58 \pm 1.44 \times 10^{43} $ & $ 2.18 \pm 0.69 \times 10^{44} $ \\ 
\hline
\end{tabular}
\end{minipage}
\end{table*}

\begin{table*}
 \centering
 \begin{minipage}{120mm}
  \caption{X--ray hard luminosities and SFR for the radio sources 
with an X--ray counterpart selected as star forming galaxies. 
X ID is from \citet{xue11} and \citet{lehmer05}; z 
is the spectroscopic or photometric redshift; $L_{2-10 keV}$ is the intrinsic X-ray luminosity
in the hard band; $L_{1.4 GHz}$ is the k corrected radio luminosity; SFRs are evaluated using 
Eq. \ref{eq:bell} \citep{bell03}. \label{tab:table1} }
 \centering
  \begin{tabular}{ccrrr}
  \hline
  \hline
  X ID & z & $L_{2-10 keV}$ ($10^{40} erg s^{-1})$ & $L_{1.4 GHz}$ ($10^{22} W Hz^{-1})$ & $SFR$ ($M_{\odot}/yr$)  \\
  \hline
  81 &   0.98 &        38.3 $\pm $   29.2 &       10.46 $\pm $   2.53 &       57.72 $\pm $  13.99 \\ 
 152 &   0.74 &        20.5 $\pm $    2.7 &        9.43 $\pm $   1.34 &       52.05 $\pm $   7.39 \\ 
 154 &   0.08 &        0.21 $\pm $   0.09 &        0.07 $\pm $   0.01 &        0.40 $\pm $   0.06 \\ 
 164 &   0.42 &        5.92 $\pm $   2.36 &        2.13 $\pm $   0.35 &       11.74 $\pm $   1.95 \\ 
 249 &   0.12 &        0.47 $\pm $   0.13 &        0.16 $\pm $   0.02 &        0.90 $\pm $   0.13 \\ 
 287 &   0.68 &        16.4 $\pm $   7.57 &        4.26 $\pm $   1.07 &       23.54 $\pm $   5.88 \\ 
 289 &   0.67 &        11.8 $\pm $   5.28 &        6.82 $\pm $   1.03 &       37.62 $\pm $   5.70 \\ 
 300 &   0.20 &        3.68 $\pm $   1.00 &        0.37 $\pm $   0.07 &        2.07 $\pm $   0.38 \\ 
 316 &   0.66 &        11.3 $\pm $   1.45 &        7.34 $\pm $   0.99 &       40.52 $\pm $   5.49 \\ 
 322 &   0.58 &        16.2 $\pm $   2.82 &        4.90 $\pm $   0.74 &       27.03 $\pm $   4.10 \\ 
 349 &   0.90 &        39.5 $\pm $   6.60 &        9.86 $\pm $   2.01 &       54.42 $\pm $  11.10 \\ 
 387 &   0.21 &        7.04 $\pm $   2.44 &        0.68 $\pm $   0.08 &        3.75 $\pm $   0.45 \\ 
 413 &   0.08 &        0.47 $\pm $   0.03 &        0.38 $\pm $   0.01 &        2.12 $\pm $   0.05 \\ 
 453 &   0.67 &        24.6 $\pm $   8.19 &        6.34 $\pm $   1.01 &       35.01 $\pm $   5.59 \\ 
 476 &   1.11 &        32.9 $\pm $   16.9 &       13.27 $\pm $   3.36 &       73.26 $\pm $  18.54 \\ 
 483 &   0.04 &        0.08 $\pm $   0.01 &        0.02 $\pm $   0.00 &        0.09 $\pm $   0.01 \\ 
 500 &   0.55 &        24.3 $\pm $   3.77 &        6.70 $\pm $   0.63 &       37.00 $\pm $   3.49 \\ 
 530 &   0.74 &        20.9 $\pm $   3.09 &       12.05 $\pm $   1.31 &       66.50 $\pm $   7.22 \\ 
 531 &   0.24 &        1.18 $\pm $   0.38 &        1.04 $\pm $   0.10 &        5.72 $\pm $   0.55 \\ 
 582 &   0.28 &        1.60 $\pm $   0.16 &        3.68 $\pm $   0.14 &       20.34 $\pm $   0.75 \\ 
 595 &   0.71 &        38.2 $\pm $   16.1 &       11.04 $\pm $   1.26 &       60.92 $\pm $   6.93 \\ 
 597 &   0.46 &        12.0 $\pm $   1.38 &        4.30 $\pm $   0.43 &       23.72 $\pm $   2.36 \\ 
 598 &   0.44 &        6.96 $\pm $   0.95 &        1.84 $\pm $   0.38 &       10.13 $\pm $   2.12 \\ 
 609 &   0.10 &        2.08 $\pm $   0.22 &        0.10 $\pm $   0.02 &        0.54 $\pm $   0.09 \\ 
 623 &   0.73 &        61.5 $\pm $   15.4 &       13.45 $\pm $   1.30 &       74.23 $\pm $   7.20 \\ 
 638 &   0.34 &        5.77 $\pm $   1.38 &        1.01 $\pm $   0.21 &        5.57 $\pm $   1.17 \\ 
 654 &   0.53 &        14.9 $\pm $   5.64 &       12.72 $\pm $   0.64 &       70.24 $\pm $   3.52 \\ 
 671 &   0.37 &        14.7 $\pm $   4.59 &        1.49 $\pm $   0.27 &        8.24 $\pm $   1.47 \\ 
 719 &   0.22 &        6.80 $\pm $   2.41 &        1.21 $\pm $   0.08 &        6.67 $\pm $   0.45 \\ 
 727 &   0.18 &        4.24 $\pm $   1.86 &        2.38 $\pm $   0.06 &       13.16 $\pm $   0.32 \\ 
 731 &   0.56 &        50.7 $\pm $   11.6 &        4.91 $\pm $   0.76 &       27.08 $\pm $   4.21 \\ 
1074 &   0.18 &        2.96 $\pm $   0.22 &        0.87 $\pm $   0.06 &        4.83 $\pm $   0.35 \\ 
1082 &   0.14 &        2.13 $\pm $   0.75 &        0.76 $\pm $   0.04 &        4.21 $\pm $   0.23 \\ 
1460 &   0.10 &        2.28 $\pm $   0.65 &        0.27 $\pm $   0.02 &        1.51 $\pm $   0.09 \\ 
1461 &   0.15 &        2.43 $\pm $   0.92 &        0.62 $\pm $   0.04 &        3.41 $\pm $   0.23 \\ 
1504 &   0.15 &        5.08 $\pm $   1.10 &        7.23 $\pm $   0.04 &       39.89 $\pm $   0.22 \\ 
1599 &   0.25 &        16.6 $\pm $   14.1 &        2.38 $\pm $   0.14 &       13.15 $\pm $   0.79 \\ 
1623 &   0.50 &        52.1 $\pm $   41.4 &        2.66 $\pm $   0.68 &       14.69 $\pm $   3.73 \\ 
1632 &   0.36 &        38.4 $\pm $   14.5 &        1.81 $\pm $   0.34 &       10.02 $\pm $   1.86 \\ 
1634 &   0.09 &        2.45 $\pm $   0.81 &        0.15 $\pm $   0.01 &        0.83 $\pm $   0.07 \\ 
1639 &   0.15 &        1.47 $\pm $   0.69 &        1.00 $\pm $   0.04 &        5.53 $\pm $   0.22 \\ 
1657 &   0.14 &        5.34 $\pm $   1.69 &        0.56 $\pm $   0.04 &        3.11 $\pm $   0.22 \\ 
1664 &   0.13 &        18.3 $\pm $   2.65 &        0.53 $\pm $   0.03 &        2.94 $\pm $   0.16 \\ 
\hline
\end{tabular}
\end{minipage}
\end{table*}

\label{lastpage}

\end{document}